\documentclass[a4paper,11pt]{article}
\usepackage{a4wide,amsmath,amssymb,bm,bbm}
\usepackage[makeroom]{cancel}
\usepackage[english]{babel}
\usepackage[utf8]{inputenc}

\usepackage{hyperref,enumerate}
\hypersetup{
    colorlinks,
    linkcolor={black},
    citecolor={black},
    urlcolor={black}
}

\makeatletter

\@addtoreset{equation}{section} \makeatother

\begin{document}
\setcounter{page}{0}

\hfill
\vspace{30pt}

\begin{center}
{\huge{\bf {Second order bosonic string effective action from $O(d,d)$}}}

\vspace{80pt}

Linus Wulff

\vspace{15pt}

\small {\it Department of Theoretical Physics and Astrophysics, Faculty of Science, Masaryk University\\ 611 37 Brno, Czech Republic}
\\
\vspace{12pt}
\texttt{wulff@physics.muni.cz}\\

\vspace{80pt}

{\bf Abstract}
\end{center}
\noindent
The corrections to the tree-level effective action for the bosonic string up to second order in $\alpha'$ are fixed by requiring its dimensional reduction to $26-d$ dimensions to be compatible with $O(d,d)$ symmetry. The result is in agreement with the literature, but takes a simpler form than previously know expressions. We identify some structures in the Lagrangian which appear at least at the first three orders in $\alpha'$.

\clearpage
\tableofcontents
\section{Introduction and summary of results}
In a quantum theory of gravity one expects the Einstein-Hilbert action, which is valid at low energies, to receive higher-derivative corrections, in the form of higher powers of the Riemann tensor multiplied by appropriate inverse powers of the Planck mass. Knowing the first few corrections allows for computing physically interesting quantities, e.g. corrections to black hole thermodynamics. In string theory the corrections to the low-energy action can be systematically computed. In this case they arise from higher loop corrections to the beta functions of the string sigma model in a general background. Alternatively, they can be found from (flat space) string scattering amplitudes. In this paper we will restrict ourselves to the simplest (but unrealistic) case of the bosonic string. We will also neglect string loop corrections, proportional to higher powers of the string coupling, $g_s=e^{\langle\Phi\rangle}$, where $\Phi$ is the dilaton.

The tree-level spacetime effective action for the bosonic string then takes the form of an infinite expansion in $\alpha'$ (inverse string tension)
\begin{equation}
S=\int d^{26}x\,\sqrt{-g}\,e^{-2\Phi}(L_0+\alpha'L_1+\alpha'^2L_2+\cdots)\,.
\label{eq:S}
\end{equation}
At lowest order we have the familiar Lagrangian for the NS-NS sector of supergravity,
\begin{equation}
L_0=R+4(\partial\Phi)^2-\frac{1}{12}H^2\,,
\end{equation}
where $H^2=H_{abc}H^{abc}$ and $H=dB$. At the first order in $\alpha'$ there is a correction involving the square of the Riemann tensor, which was first computed by Metsaev and Tseytlin in \cite{Metsaev:1987bc,Metsaev:1987zx}. They also computed the terms involving the metric at order $\alpha'^2$ in \cite{Metsaev:1986yb}, which turn out to involve terms cubic in the Riemann tensor. The full Lagrangian at order $\alpha'^2$, i.e. including also all dilaton and $B$-field terms, was only found rather recently by Garousi \cite{Garousi:2019mca}. 

Rather than using string amplitude calculations, which would require going up to six-point amplitudes, it was computed using T-duality. The approach is the following. One starts from a general ansatz for the effective action at order $\alpha'$ in 26 dimensions \cite{Garousi:2019cdn}. One then reduces on a circle and requires the result to be invariant under T-duality \cite{Garousi:2019mca}. The result was recently simplified and shown to match with string amplitude calculations at four points \cite{Gholian:2023kjj}. While this is a powerful technique, the main disadvantage of this approach is that it has to be carried out with the aid of a computer, due to the large number of terms in a general ansatz for the effective action. For the same reason it usually leads to a complicated result, which must then be simplified by hand.

Here we find a result that agrees with \cite{Gholian:2023kjj} at order $\alpha'^2$, but is actually simpler at order $\alpha'$ (the two results are of course related by field redefinitions). The approach we use is also closely related to T-duality, but it avoids the need for taking a complete ansatz in $D=26$ and reducing it on a circle. Instead, while the calculations are still long, it is possible to carry them out without the aid of a computer. In this way we are able to identify certain structures in the result which are otherwise very hard to recognize.

The idea is to use the fact that the tree-level string effective action reduced to $D-d$ dimensions has a (continuous) $O(d,d)$ symmetry \cite{Meissner:1991zj,Meissner:1991ge,Sen:1991zi} (see also \cite{Maharana:1992my}). The requirement that the reduced theory have this symmetry places very strong constraints on the unreduced $D$-dimensional theory. Of course, starting from a general ansatz in $D=26$, reducing it and requiring $O(d,d)$ symmetry is even more complicated than the calculations of \cite{Garousi:2019mca}, which had to be carried out with the help of a computer, so this approach does not seem promising. Instead we will simplify the problem, by only imposing a certain \emph{necessary} condition for $O(d,d)$ symmetry. For this it will be enough to consider only the two KK vector fields that arise in the dimensional reduction, one coming from the metric and the other from the $B$-field, while we ignore all terms involving the KK scalars. The two sets of $d$ KK vectors combine to a single $O(d,d)$ vector. In the reduced action any internal indices must be contracted, so looking at the quadratic terms in the KK vectors, there are three possible structures that can arise $A_m\cdot A_n$, $\hat A_m\cdot\hat A_n$ and $A_m\cdot\hat A_n$, where $A_m^{n'}$ and $\hat A_m^{n'}$ is the sum/difference of the two KK vectors and we used a `dot' to denote the contraction of an internal (primed) index (actually, because of gauge invariance only the field strengths, rather than the vectors themselves, appear in the action). We have three possible contractions, but there are only \emph{two} possible $O(d,d)$ invariants, formed by contracting the $O(d,d)$ vector with either the constant $O(d,d)$ invariant metric or with the so-called generalized metric (constructed from the metric and $B$-field). It turns out that the contractions $A_m\cdot A_n$ and $\hat A_m\cdot\hat A_n$ are compatible with the $O(d,d)$ symmetry, while the contraction $A_m\cdot\hat A_n$ explicitly violates it. A necessary condition for $O(d,d)$ invariance is then that all terms involving the $O(d,d)$ violating contraction cancel out, up to total derivatives and terms proportional to the equations of motion, since the latter can be removed by field redefinitions. This condition turns out to be extremely strong and fixes the form of the 26-dimensional Lagrangian almost completely.\footnote{There is a small ambiguity in that terms involving only derivatives of the dilaton are not fixed. Such terms are not expected to arise (except at zeroth order in $\alpha'$) and indeed they are known not to be present up to $\alpha'^3$, so this is not a big problem.} 

This approach was introduced in \cite{Wulff:2021fhr}, where it was used to find all terms involving the metric and $B$-field for the Riemann$^4$ correction at order $\alpha'^3$ up to fifth order in fields. Here we will use it to find the full tree-level bosonic string effective action up to order $\alpha'^2$. Note that our goal is only to constrain the form of the original $D$-dimensional effective action, not to demonstrate the $O(d,d)$ symmetry of the dimensionally reduced action. The latter is much more involved since in that case we cannot ignore the terms involving the KK scalars. However, it is not necessary for us to do this as the $O(d,d)$ symmetry (after field redefinitions) is already guaranteed by the general argument in \cite{Sen:1991zi}.

The reason we focus on the bosonic string here, rather than the more interesting supersymmetric string theories, is that the corrections are simpler, while the field content and symmetry under $B\rightarrow-B$ is the same as for the NSNS sector of the type II string. The hope is therefore to find some structures in the answer that come from $O(d,d)$ symmetry and occur at several different orders in $\alpha'$ and both for the bosonic string and the type II string. Indeed, this is what we find for many of the terms, although others remain somewhat mysterious. It is also for this reason that we don't use the manifestly $O(d,d)$ invariant formulation known as Double Field Theory (DFT). While the first and second $\alpha'$ correction to the bosonic string can be described in this formalism \cite{Hohm:2014xsa,Marques:2015vua,Baron:2018lve,Baron:2020xel,Hronek:2021nqk,Hronek:2022dyr}, it appears to fail for the $\alpha'^3$ correction \cite{Hronek:2020xxi}, so it would not be useful for seeing what is common to all these corrections.

Since the end result will be of more interest to most readers than the derivation we present it here. At the first order in $\alpha'$ we find the correction
\begin{equation}
\begin{split}
L_1
&=
t_4t_4\mathcal R^2
-t_4t_4H^2\mathcal R
-\frac13H^{abc}H^3_{abc}
\\
&=
\frac12t_4t_4\mathcal R^2
-\frac12t_4t_4H^2\mathcal R
-\frac18(\varepsilon_4\varepsilon_4)'\mathcal R^2
+\frac{1}{72}(\varepsilon_5\varepsilon_5)'H^2\mathcal R
\\
&\quad
-\frac13\nabla^eH^{abc}\nabla_eH_{abc}
-\frac13H^{abc}H^3_{abc}\,.
%
\end{split}
\end{equation}
This result is well known and agrees with \cite{Metsaev:1987bc} (sending $\alpha'\rightarrow\frac14\alpha'$ in (\ref{eq:S})). Its $O(d,d)$ symmetry was discussed in \cite{Meissner:1996sa,Eloy:2019hnl}. In the first line we have written the standard result, using a compact notation, while in the second we have rewritten it in a way that is very similar to expressions that appear at higher orders in $\alpha'$. Here $\mathcal R$ is the curvature constructed from the torsionful spin connection 
\begin{equation}
\omega^{(-)}=\omega-\frac12H\,,
\end{equation}
which takes the form
\begin{equation}
\mathcal R_{abcd}=R_{abcd}-(\nabla H)_{abcd}+\frac12H^2_{a[cd]b}\,,
\label{eq:calR}
\end{equation}
where we defined $(\nabla H)_{abcd}=\nabla_{[a}H_{b]cd}$ and $H^2_{abcd}=H_{abe}H^e{}_{cd}$. The structures appearing in the Lagrangian are defined as follows
\begin{equation}
t_4t_4\mathcal R^2=\mathcal R^{ab}{}_{cd}\mathcal R_{ab}{}^{cd}\,,
\end{equation}
i.e. a trace over the first two and last two indices. Similarly $t_4t_4H^2\mathcal R=H^2_{abcd}\mathcal R^{abcd}$. We have also defined
\begin{equation}
(\varepsilon_4\varepsilon_4)'\mathcal R^2=-4!(\mathcal R^{ab}{}_{[ab}\mathcal R^{cd}{}_{cd]})'=-4\mathcal R^{ab}{}_{cd}\mathcal R^{cd}{}_{ab}
\end{equation}
and
\begin{equation}
(\varepsilon_5\varepsilon_5)'H^2\mathcal R=-5!(H^{abc}H_{[abc}\mathcal R^{de}{}_{de]})'=-36H^2_{abcd}\mathcal R^{abcd}\,,
\end{equation}
which are not independent at this order but have direct analogs appearing at higher order in $\alpha'$. Note that the prime in these expressions denotes removal of terms with self-contractions, i.e. torsionful Ricci tensor/scalar terms. Finally, we have defined
\begin{equation}
H^3_{abc}=H_{ade}H_b{}^{ef}H_{cf}{}^d\,,
\label{eq:H3}
\end{equation}
i.e. with a trace over the last two indices, which is completely antisymmetric in $a,b,c$.

Our new result is the $\alpha'^2$ correction which takes the form
\begin{equation}
\begin{split}
L_2
&=
\frac23t_6t_6\mathcal R^3
-t_6t_6H^2\mathcal R^2
+2t_6t_6H^2(\nabla H)^2
-2t_6t_6(H^2)^2\mathcal R
\\
&\quad
-\frac{1}{32}(\varepsilon_6\varepsilon_6)'\mathcal R^3
-\frac{1}{192}(\widehat{\varepsilon_7\varepsilon_7})'H^2\mathcal R^2
+\frac{1}{96}(\widehat{\varepsilon_7\varepsilon_7})'H^2(\nabla H)^2
\\
&\quad
+\frac{4!}{16}H^2_{[abcd]}\left((\nabla H)^{abef}(\nabla H)^{cd}{}_{ef}-\nabla^eH^{fab}\nabla_eH_f{}^{cd}\right)
\\
&\quad
-\frac12H_{ab[e}\nabla^gH^{ab}{}_{f]}H^{cde}\nabla_gH_{cd}{}^f
+\frac12\nabla_cH^2_{ab}\nabla^{[a}(H^2)^{c]b}
+H^{abe}(\nabla H)_{ab}{}^{cd}\nabla_cH^2_{de}
\\
&\quad
+\frac16H^2_{ab}\nabla^aH^{def}\nabla^bH_{def}
-\frac13H^3_{abc}(H^3)^{abc}\,,
\end{split}
\label{eq:L2}
\end{equation}
where we have defined
\begin{equation}
t_6t_6\mathcal R^3=\mathcal R^{ab}{}_{ef}\mathcal R_{bc}{}^{fg}\mathcal R^c{}_{ag}{}^e\,,
\label{eq:t6}
\end{equation}
i.e. again involving a trace over the first two and last two indices. The other $t_6t_6$ terms are defined similarly, e.g. $t_6t_6(H^2)^2\mathcal R=(H^2)^{ab}{}_{ef}(H^2)_{bc}{}^{fg}\mathcal R^c{}_{ag}{}^e$. We have also defined
\begin{equation}
(\varepsilon_6\varepsilon_6)'\mathcal R^3=-6!(\mathcal R^{ab}{}_{[ab}\mathcal R^{cd}{}_{cd}\mathcal R^{ef}{}_{ef]})'
\label{eq:e6}
\end{equation}
and
\begin{equation}
(\widehat{\varepsilon_7\varepsilon_7})'H^2\mathcal R^2=(\varepsilon_7\varepsilon_7)'H^2\mathcal R^2-(\varepsilon_7\varepsilon_7)'[H^2]\mathcal R^2-6H^2(\varepsilon_4\varepsilon_4)'\mathcal R^2\,,
\label{eq:e7hat}
\end{equation}
where $H^2=H_{abc}H^{abc}$ and
\begin{equation}
(\varepsilon_7\varepsilon_7)'H^2\mathcal R^2=-7!(H^{abc}H_{[abc}\mathcal R^{de}{}_{de}\mathcal R^{fg}{}_{fg]})'\,,
\label{eq:e7}
\end{equation}
while
\begin{equation}
(\varepsilon_7\varepsilon_7)'[H^2]\mathcal R^2=(4!)^2H^{abc}H_{def}(\mathcal R^{[de}{}_{[ab}\mathcal R^{fg]}{}_{cg]})'\,,
\end{equation}
which is precisely the contribution from the previous expression with no contraction between the two $H$s, i.e. $(\widehat{\varepsilon_7\varepsilon_7})'H^2\mathcal R^2$ is just $(\varepsilon_7\varepsilon_7)'H^2\mathcal R^2$ minus the terms with either all indices or no indices on the $H$s contracted. The remaining terms have a less familiar structure. Instead we have, in the third line, two terms involving the completely antisymmetrized $H^2_{[abcd]}$. Curiously, similar terms appear also at order $\alpha'^3$ \cite{Wulff:2024mgu}. The remaining quartic terms involve either a `Chern-Simons like' combination $H_{ab[c}\nabla^eH_{d]}{}^{ab}$ or $H^2_{ab}=H_{acd}H_b{}^{cd}$, while the last term has the form $H^6$.

Note that $O(d,d)$ fixes the relative coefficient compared to $L_1$, so this is not an independent correction as far as $O(d,d)$ is concerned. The above result is the most compact form known to date for the $\alpha'^2$ correction. The metric terms agree with \cite{Metsaev:1986yb} and in fact the whole expression agrees with the expression found recently in \cite{Gholian:2023kjj} (up to field redefinitions), which was shown to agree with amplitude calculations up to four points. However, the simple form for the $\alpha'^2$ Lagrangian was found in \cite{Gholian:2023kjj} only in the case where the first order Lagrangian takes a more complicated form, referred to there as the `Meissner scheme'. Our results show that this simplification occurs also in the `Metsaev-Tseytlin scheme' considered there, up to a simple field redefinition at order $\alpha'$, making the total action up to order $\alpha'^2$ much simpler. Although we don't fully understand why $O(d,d)$ fixes this precise form for the effective action, we see that many terms take a similar form at order $\alpha'$ and $\alpha'^2$ and, it turns out, also at order $\alpha'^3$, while other terms don't seem to have an obvious analog at other orders in $\alpha'$.

The remainder of this paper is devoted to the derivation of these results. In Sec. \ref{sec:KK} we review the Kaluza-Klein reduction procedure focusing on the terms that involve the KK vectors and which don't respect $O(d,d)$. In Sec. \ref{sec:R} we consider, as a warm-up, the action at zeroth order in $\alpha'$ and show that the $O(d,d)$ violating terms cancel out upon reduction. Next we consider the first $\alpha'$ correction in Sec. \ref{sec:R2} and show how cancellation of the $O(d,d)$ violating terms fixes the form of the Lagrangian and derive the field redefinitions needed to achieve $O(d,d)$ invariance. The calculation at order $\alpha'^2$ is carried out in Sec. \ref{sec:R3} and the result is then simplified in Sec. \ref{sec:simplify}. Some details of the calculation, such as the reduction of the $\varepsilon$-terms and the form of the terms proportional to the equations of motion (from which the required field redefinitions can be read off directly), are relegated to the appendix.

\section{Kaluza-Klein reduction and \texorpdfstring{$O(d,d)$}{O(d,d)}}\label{sec:KK}
A nice discussion of Kaluza-Klein reduction and $O(d,d)$ symmetry can be found in \cite{Maharana:1992my}. Here we will summarize what we need for our purposes, in particular how $O(d,d)$ violating contractions of two KK vector fields can appear. We start with the usual KK ansatz for the vielbein
\begin{equation}
\underline e_{\underline m}{}^{\underline a}
=
\left(
\begin{array}{cc}
	e_m{}^a & A^{(1)n'}_me_{n'}{}^{a'}\\
	0 & e_{m'}{}^{a'}
\end{array}
\right)\,,
\end{equation}
and the $B$-field
\begin{equation}
\underline B_{\underline {mn}}
=
\left(
\begin{array}{cc}
	\hat B_{mn} & A^{(2)}_{mn'}+A^{(1)k'}_mB_{k'n'}\\
	-A^{(2)}_{nm'}-A^{(1)k'}_nB_{k'm'} & B_{m'n'}
\end{array}
\right)\,,
\end{equation}
where
\begin{equation}
\hat B_{mn}=B_{mn}-A_{[m}^{(1)m'}A_{n]m'}^{(2)}+A^{(1)k'}_mA^{(1)l'}_nB_{k'l'}\,.
\end{equation}
Finally the reduction of the dilaton gives
\begin{equation}
\underline\Phi=\Phi+\frac12\log\det g_{m'n'}\,,
\end{equation}
where $g_{m'n'}=e_{m'}{}^{a'}e_{n'}{}^{b'}\delta_{a'b'}$.

Besides the lower-dimensional metric, $B$-field and dilaton we obtain $2d$ vector fields $A^{(1)m'}$ and $A_{m'}^{(2)}$ (we suppress, for the moment, the external vector index) as well as a number of scalar fields contained in $g_{m'n'}$ and $B_{m'n'}$. These can be combined into fields that transform simply under the $O(d,d)$ symmetry: an $O(d,d)$ vector
\begin{equation}
\mathcal A_M=
\left(
\begin{array}{c}
	A^{(1)m'}\\
	A^{(2)}_{m'}
\end{array}
\right)
\end{equation}
and a symmetric $O(d,d)$ matrix, sometimes called the generalized metric,
\begin{equation}
\mathcal H_{MN}=
\left(
\begin{array}{cc}
	g^{m'n'} & -g^{m'k'}B_{k'n'}\\
	B_{m'k'}g^{k'n'} & g_{m'n'}-B_{m'k'}g^{k'l'}B_{l'n'}
\end{array}
\right)\,.
\end{equation}
An important fact is that from two $\mathcal A$'s we can form precisely two $O(d,d)$ scalars, namely 
\begin{equation}
\mathcal A_M\eta^{MN}\mathcal A_N\qquad\mbox{and}\qquad\mathcal A_M\mathcal H^{MN}\mathcal A_N\,,
\end{equation}
where $\mathcal H^{MN}=\eta^{MK}\mathcal H_{KL}\eta^{LN}$ and we introduced the $O(d,d)$ invariant metric
\begin{equation}
\eta^{MN}=
\left(
\begin{array}{cc}
0 & \delta^{m'}_{n'}\\
\delta_{m'}{}^{n'} & 0	
\end{array}
\right)\,.
\end{equation}

For our purposes it will actually be enough to consider only the vectors and simply set the scalars to zero by imposing\footnote{Note that this is not consistent in general since the equations of motion for the scalars will contain terms not involving the scalars. However, in our case this is not a problem because the source terms in question must be quadratic in the vectors (since the scalars have two internal indices). Therefore, terms involving the equations of motion of the scalars in the Lagrangian can only mix with terms involving four vector fields, since the two free internal indices must be contracted with something, but in our calculations we will only consider terms in the Lagrangian quadratic in the vectors. The only possible exception is the trace of the scalar equation of motion, which could give something quadratic in the vectors, but since the equations of motion respect the $O(d,d)$ symmetry this cannot contribute to the $O(d,d)$ violating terms we are interested in.
}
\begin{equation}
e_{m'}{}^{a'}=\delta_{m'}^{a'}\,,\qquad B_{m'n'}=0\,.
\end{equation}
The two $O(d,d)$ scalars that can be formed from two $\mathcal A$s become, in this case, simply (internal indices are now raised and lowered with the Kronecker delta)
\begin{equation}
A_{(m}^{(1)m'}A^{(2)}_{n)m'}\qquad\mbox{and}\qquad A_m^{(1)m'}A^{(1)}_{nm'}+A_m^{(2)m'}A^{(2)}_{nm'}\,.
\end{equation}
It is convenient to define the combinations\footnote{The sign of $A$ is chosen to agree with the conventions of \cite{Wulff:2021fhr}.}
\begin{equation}
A=-\frac12(A^{(1)}+A^{(2)})\qquad\mbox{and}\qquad\hat A=\frac12(A^{(1)}-A^{(2)})\,.
\end{equation}
In terms of these the two $O(d,d)$ invariants are
\begin{equation}
A_m\cdot A_n\qquad\mbox{and}\qquad\hat A_m\cdot\hat A_n\,,
\end{equation}
where the $\cdot$ denotes contraction of the internal index. Conversely, the inner product between the two
\begin{equation}
A_m\cdot\hat A_n
\end{equation}
explicitly \emph{violates} the $O(d,d)$ symmetry. Such terms arise in the dimensional reduction, but they have to cancel out in the reduced action since otherwise $O(d,d)$ symmetry would be violated (up to total derivatives and equation of motion terms). This observation gives a powerful way to constrain the form of the effective action by requiring all such terms to cancel out.

We will need the reduction of $H$, the spin connection and Riemann tensor, as these appear in the effective action. Setting the scalars to zero simplifies things a lot and we find for $H$
\begin{equation}
\begin{aligned}
\underline H_{a'b'c'}&=0\,,& & & & & \underline H_{abc'}&=-F_{c'ab}-\hat F_{c'ab}\,,\\
\underline H_{ab'c'}&=0\,, & & & & & \underline H_{abc}&=\hat H_{abc}\,,
\end{aligned}
\label{eq:Hred}
\end{equation}
where $\hat H_{abc}=H_{abc}+3(\hat A_{[a}\cdot \hat F_{bc]}-A_{[a}\cdot F_{bc]})$ whose form we will not need since the contraction of internal indices respects the $O(d,d)$ symmetry. For the spin connection we find
\begin{equation}
\begin{aligned}
%
\underline\omega_{a'b'c'}&=0\,,& & & & & \underline\omega_{a'bc}&=\frac12(F_{a'bc}-\hat F_{a'bc})\,,\\
\underline\omega_{a'b'c}&=0\,,& & & & & \underline\omega_{abc'}&=\frac12(-F_{c'ab}+\hat F_{c'ab})\,,\\
\underline\omega_{ab'c'}&=0\,,& & & & & \underline\omega_{abc}&=\omega_{abc}\,.
%
\end{aligned}
\end{equation}
Using this it is not hard to compute the reduction of the Riemann tensor. From the perspective of $O(d,d)$ symmetry it is actually easier to work with the torsionful Riemann tensor defined in (\ref{eq:calR}). A short calculation gives the KK reduction\footnote{To go from the conventions of \cite{Wulff:2021fhr} to those of the present paper one sets
$$
F_{a',\mathrm{there}}=\frac{1}{\sqrt2}\hat F_{a',\mathrm{here}}\,,\qquad F^{a'}_{\mathrm{there}}=\frac{1}{\sqrt2}F^{a'}_{\mathrm{here}}\,.
$$
}
\begin{equation}
\begin{aligned}
\underline{\mathcal R}_{a'b'c'd'}&=0\,,& & & & & \underline{\mathcal R}_{abc'd'}&=-2\hat F_{c'[a}{}^e\hat F_{d'b]e}\,,& & & & & \underline{\mathcal R}_{abc'd}&=\nabla^{(+)}_d\hat F_{c'ab}\,,\\
\underline{\mathcal R}_{a'b'c'd}&=0\,,& & & & & \underline{\mathcal R}_{a'b'cd}&=-2F_{a'[c}{}^eF_{b'd]e}\,,& & & & & \underline{\mathcal R}_{a'bcd}&=-\nabla^{(-)}_bF_{a'cd}\,,\\
\underline{\mathcal R}_{ab'c'd'}&=0\,,& & & & & \underline{\mathcal R}_{a'bc'd}&=\hat F_{c'b}{}^eF_{a'ed}\,,& & & & & \underline{\mathcal R}_{abcd}&=\hat{\mathcal R}_{abcd}+\hat F_{ab}\cdot F_{cd}\,,
\end{aligned}
\label{eq:Rred}
\end{equation}
where we defined $\hat{\mathcal R}_{abcd}=\mathcal R_{abcd}+2\hat F_{a[c}\cdot\hat F_{d]b}-F_{ab}\cdot F_{cd}$, whose form we will again not need since the contraction of internal indices respects the $O(d,d)$ symmetry. In contrast, the second term in the last expression violates the $O(d,d)$ symmetry and will therefore be important for us. In the above expressions it is understood that $H$ is everywhere replaced by $\hat H$ defined below (\ref{eq:Hred}), e.g. $\nabla^{(\pm)}=\nabla\pm\frac12\hat H$. For completeness we also compute the reduction of the torsionful Ricci tensor and scalar
\begin{equation}
\begin{aligned}
\underline{\mathcal R}_{a'b'}=&\,{}-F_{a'cd}\hat F_{b'}^{cd}\\
\underline{\mathcal R}_{ab'}=&\,\nabla^{(+)c}\hat F_{b'ac}\\
\underline{\mathcal R}_{a'b}=&\,{}-\nabla^{(-)c}F_{a'bc}\\
\underline{\mathcal R}_{ab}=&\,\hat{\mathcal R}_{ab}
\end{aligned}
\label{eq:Ricci-red}
\end{equation}
and
\begin{equation}
\underline{\mathcal R}=\hat{\mathcal R}-\hat F_{ab}\cdot F^{ab}\,,
\label{eq:RicciS-red}
\end{equation}
with $\hat{\mathcal R}_{ab}$ and $\hat{\mathcal R}$ defined by contracting $\hat{\mathcal R}_{abcd}$ in the usual way.

Actually, since we will restrict ourselves to terms in the reduced action which are quadratic in the KK field strengths and violate $O(d,d)$, i.e. the terms involving $F\cdot\hat F$, we will not need any quantities involving two free internal indices, as these can only contribute to terms of fourth power in the KK field strengths.

\section{Zeroth order: Ricci scalar}\label{sec:R}
At the lowest order in $\alpha'$ the tree-level bosonic string effective action is
\begin{equation}
S_0=\int d^{26}x\,\sqrt{-g}\,e^{-2\Phi}L_0\,,
\end{equation}
with the Lagrangian
\begin{equation}
L_0=R+4(\partial\Phi)^2-\frac{1}{12}H^2\,,
\end{equation}
where $H^2=H_{abc}H^{abc}$. Using the torsionful Riemann tensor (\ref{eq:calR}) we can write it instead as
\begin{equation}
L_0=\mathcal R+4(\partial\Phi)^2+\frac16H^2\,.
\end{equation}
Carrying out the dimensional reduction, displaying only the $O(d,d)$ violating terms involving $F\cdot\hat F$ we have, using (\ref{eq:RicciS-red}) and (\ref{eq:Hred}),
\begin{equation}
\mathcal R\rightarrow-\hat F_{ab}\cdot F^{ab}\,,\qquad H^2\rightarrow 6\hat F_{ab}\cdot F^{ab}\,,
\end{equation}
while the dilaton term gives no contribution. We see that the coefficient of $H^2$ in $L_0$ is precisely the one required for these $O(d,d)$ violating terms to cancel.

We will now use the requirement that the $O(d,d)$ violating terms cancel out to fix the form of the effective action at order $\alpha'$ and $\alpha'^2$ starting from the knowledge of the terms involving only the metric, i.e. the Riemann tensor squared and cubed respectively. Note that as above some terms involving only derivatives of the dilaton are not fixed by our procedure, specifically terms built only from $\partial\Phi$ or $\partial^a\Phi\nabla_a\partial_b\Phi$ as these do not give rise to a term involving $F\cdot\hat F$ upon dimensional reduction. To fix also these terms one would need to include also the scalars from the dimensional reduction, resulting in much longer calculations. However, these terms are very special, and not expected to arise in the effective action. Indeed, setting these terms to zero we find an effective action without terms involving the dilaton which is in agreement with known results in the literature at order $\alpha'$ and $\alpha'^2$.

\section{First order: Riemann squared correction}\label{sec:R2}
At the first order in $\alpha'$ the tree-level bosonic string effective action is known to have a correction involving the square of the Riemann tensor
\begin{equation}
L_1=R_{abcd}R^{abcd}+\ldots\,,
\end{equation}
where we set the coefficient in front (which is not fixed by $O(d,d)$) to $1$ for simplicity. We will now determine the remaining terms by requiring the $O(d,d)$ violating $F\cdot\hat F$ terms to vanish. First we should replace $R_{abcd}$ by the torsionful Riemann tensor $\mathcal R_{abcd}$ defined in (\ref{eq:calR}). Using (\ref{eq:Rred}) one finds that there is only one possible contraction of two of these which does not violate $O(d,d)$ already at the quadratic order in fields, namely
\begin{equation}
\mathcal R_{abcd}\mathcal R^{abcd}\,,
\end{equation}
which gives\footnote{Recall that on the RHS we keep only the $F\cdot\hat F$ terms from the dimensional reduction, in particular $O(d,d)$ violating terms involving $(F\cdot\hat F)^2$ are not considered.}
\begin{equation}
\mathcal R_{abcd}\mathcal R^{abcd}\rightarrow 2\hat F_{ab}\cdot F_{cd}\mathcal R^{abcd}\,.
\label{eq:R2red}
\end{equation}
It is clear that this can only be canceled by including $H^2\mathcal R$ terms. Noting that the bosonic string effective action can only contain even powers of $H$, due to the symmetry under $B\rightarrow-B$, there is only one possible term (we will not include a term involving the Ricci tensor as it could be removed by a field redefinition) and using (\ref{eq:Hred}) and (\ref{eq:Rred}) we have
\begin{equation}
\begin{split}
H^2_{abcd}\mathcal R^{abcd}
&\rightarrow
(\hat F_{ab}\cdot F_{cd}+F_{ab}\cdot\hat F_{cd})\mathcal R^{abcd}
-2\hat F^{eb}\cdot\nabla^{(-)}_bF^{cd}H_{ecd}
+2F^{ec}\cdot\nabla^{(+)}_c\hat F^{ab}H_{abe}
\\
&\quad
+\hat F^{ab}\cdot F^{cd}H^2_{abcd}\,.
\end{split}
\end{equation}
We need to first rewrite the third and fourth term to get rid of the derivatives acting on $F$ and $\hat F$. There are actually two ways to do this. We could integrate by parts to get\footnote{Note that  since the measure contains the dilaton factor $e^{-2\Phi}$ we pick up a term $-2\partial\Phi$ whenever we integrate by parts. We will always drop total derivative terms in the effective action.}
\begin{equation}
\begin{split}
&
2(\nabla^{(+)}_b-2\partial_b\Phi)\hat F^{eb}\cdot F^{cd}H_{ecd}
-2(\nabla^{(-)}_c-2\partial_c\Phi)F^{ec}\cdot\hat F^{ab}H_{abe}
-4\hat F_{ab}\cdot F_{cd}(\nabla H)^{abcd}
\\
&
-2\hat F^{ab}\cdot F^{cd}H^2_{abcd}
-4\hat F^{ab}\cdot F^{cd}H^2_{acbd}\,.
\end{split}
\end{equation}
The first two terms are proportional to the lowest order equations of motion for $\hat A$ and $A$. We will come back to these terms. On the other hand we could first use the Bianchi identity which gives instead
\begin{equation}
-4\nabla_c(\hat F_{eb}\cdot F^b{}_d)H^{ecd}
-4\hat F^{ab}\cdot F^{cd}H^2_{acbd}\,.
\end{equation}
After integration by parts the first term becomes proportional to the equation of motion for $B$. Combining these two ways of rewriting the terms we find
\begin{equation}
\begin{split}
H^2_{abcd}\mathcal R^{abcd}
&\rightarrow
2\hat F_{ab}\cdot F_{cd}\mathcal R^{abcd}
-4\hat F^{ab}\cdot F^{cd}H^2_{acbd}
+2\hat F_{eb}\cdot F^b{}_d(\nabla_c-2\partial_c\Phi)H^{ecd}
\\
&\quad
+(\nabla^{(+)}_b-2\partial_b\Phi)\hat F^{eb}\cdot F^{cd}H_{ecd}
-(\nabla^{(-)}_c-2\partial_c\Phi)F^{ec}\cdot\hat F^{ab}H_{abe}\,,
\end{split}
\end{equation}
where, to get the first term, we used the fact that we can split $\mathcal R_{abcd}$ into a piece symmetric under exchanging the two pairs of indices and one anti-symmetric,
\begin{equation}
\mathcal R^{abcd}=\mathcal R_S^{abcd}-(\nabla H)^{abcd}=\mathcal R_S^{cdab}+(\nabla H)^{cdab}\,,\qquad \mathcal R_S^{abcd}=R^{abcd}+\frac12(H^2)^{a[cd]b}\,.
\end{equation}
The first term is now of the form of the term we needed to cancel in (\ref{eq:R2red}). The last three terms involve equations of motion and we will come back to these. This leaves the second term which should be canceled. This can indeed be done by adding an $H^4$ term of the form
\begin{equation}
H^2_{abcd}(H^2)^{acbd}\rightarrow 12\hat F^{ab}\cdot F^{cd}H^2_{acbd}\,.
\end{equation}
Putting these results together we see that the Lagrangian\footnote{Note that $H^2_{abcd}(H^2)^{acbd}=H_{abc}(H^3)^{abc}$ from (\ref{eq:H3}).}
\begin{equation}
L_1=\mathcal R_{abcd}\mathcal R^{abcd}
-H^2_{abcd}\mathcal R^{abcd}
-\frac13H_{abc}(H^3)^{abc}
\label{eq:L1}
\end{equation}
only gives rise to $O(d,d)$ violating terms involving the equations of motion. This is actually fine, because such terms can be canceled by field redefinitions in the reduced theory. Therefore we have succeeded in completing the Riemann squared term to something compatible with $O(d,d)$ symmetry upon dimensional reduction. Indeed, our result matches the known form of the first $\alpha'$ correction \cite{Metsaev:1987bc,Metsaev:1987zx} for the bosonic string (up to the overall coefficient which is not fixed in our approach). When expressed in terms of the standard Riemann tensor it takes the form (dropping again total derivatives)
\begin{equation}
L_1
=
R_{abcd}R^{abcd}
-\frac12H^2_{abcd}R^{abcd}%
-\frac18H^2_{ab}(H^2)^{ab}
+\frac{1}{24}H_{abc}(H^3)^{abc}
+\mathbbm B_{ab}\mathbbm B^{ab}
-H^2_{ab}\mathbbm G^{ab}\,,
\label{eq:L1ord}
\end{equation}
where the last two terms can be removed by field redefinitions since they are proportional to the equations of motion for the $B$-field and metric,
\begin{equation}
\mathbbm B_{ab}=(\nabla^c-2\partial^c\Phi)H_{abc}\,,\qquad
\mathbbm G_{ab}=R_{ab}+2\nabla_a\partial_b\Phi-\frac14H^2_{ab}\,.
\label{eq:eom1}
\end{equation}
The derivation presented here is by far the quickest way we are aware of to arrive at this correction. For example, it is far simpler than closely related approaches which also use $O(d,d)$, e.g. \cite{Godazgar:2013bja,Baron:2022but,David:2022jcl}.

If we were only interested in the correction at order $\alpha'$ we would be done. However, we are interested also in the next order, $\alpha'^2$, and for this reason we need to look in detail at the field redefinitions needed to remove the $O(d,d)$ violating terms at order $\alpha'$, because they will generate terms of order $\alpha'^2$. Our calculations have shown that 
\begin{equation}
L_1\rightarrow
\hat{\mathbbm A}^e\cdot F^{cd}H_{ecd}
-\mathbbm A^e\cdot\hat F^{ab}H_{abe}
-2\hat F_{eb}\cdot F^b{}_d\mathbbm B^{de}\,,
\end{equation}
where the equations of motion for the KK vectors are (recall that primed indices are internal)
\begin{equation}
\mathbbm A^b_{a'}=(\nabla^{(-)}_a-2\partial_a\Phi)F^{ab}_{a'}\,,\qquad \hat{\mathbbm A}^b_{a'}=(\nabla^{(+)}_a-2\partial_a\Phi)\hat F^{ab}_{a'}\,.
\label{eq:eom2}
\end{equation}
Note that we denote the equations of motion for each field (except the dilaton, whose equation of motion will not enter) by the corresponding blackboard boldface letter, $\mathbbm G$, $\mathbbm B$, $\mathbbm A$ and $\hat{\mathbbm A}$.

The first two terms can be canceled by a field redefinition of the KK vector fields $A$ and $\hat A$, but it will be more convenient for us to trade this for a redefinition of the metric and $B$-field. The reason is that this avoids having to deal with the full set of terms quadratic in the KK vectors in the reduced action (which we did not determine since we only kept the $O(d,d)$ violating terms).  To do this we first note that, using the fact that 
\begin{equation}
\mathcal R_{ab}+2\nabla^{(-)}_a\partial_b\Phi=\mathbbm G_{ab}+\frac12\mathbbm B_{ab}\,,
\end{equation}
together with (\ref{eq:Ricci-red}), we have
\begin{equation}
H^2_{ab}\mathbbm G^{ab}\rightarrow
\hat{\mathbbm A}^e\cdot F^{cd}H_{ecd}
-\mathbbm A^e\cdot\hat F^{ab}H_{abe}
-4F_a{}^c\cdot\hat F_{cb}\mathbbm G^{ab}\,.
\end{equation}
Defining the new Lagrangian
\begin{equation}
L'_1=L_1-H^2_{ab}\mathbbm G^{ab}\,,
\label{eq:L1prime}
\end{equation}
which is related to $L_1$ by a field redefinition, we find
\begin{equation}
L'_1
\rightarrow
4F_a{}^c\cdot\hat F_{cb}\mathbbm G^{ab}
-2\hat F_{eb}\cdot F^b{}_d\mathbbm B^{de}\,,
\end{equation}
which now involves only the equations of motion for the metric and $B$-field (of the reduced theory) on the RHS. These contributions can now be canceled by the following field redefinitions (in the reduced theory)\footnote{This can be lifted to a non-covariant field redefinition in the 26-dimensional theory by observing that
$$
-2\omega^{(-)cd}_a\omega^{(+)}_{bcd}(\mathcal R^{ab}+2\nabla^{(-)a}\partial^b\Phi)
\rightarrow
-4\hat F^{ac}\cdot F_c{}^b(\mathbbm G_{ab}+\frac12\mathbbm B_{ab})\,.
$$
This agrees with the field redefinition one finds from DFT \cite{Marques:2015vua}. 
}
\begin{equation}
\delta g_{mn}=\Delta_{(mn)}\,,\qquad
\delta B_{mn}=\Delta_{[mn]}\,,\qquad
\delta\Phi=\frac14\Delta_m{}^m\,,
%
\end{equation}
where we defined
\begin{equation}
\Delta_{mn}=4F_m{}^k\cdot\hat F_{kn}\,.
\label{eq:Delta}
\end{equation}
While these field redefinitions remove the remaining $O(d,d)$ violating terms at order $\alpha'$ they will also generate terms of higher order in $\alpha'$. Since we want to work out the effective action to order $\alpha'^2$ we must take into account the terms generated at this order by the field redefinition.

\subsection{Terms generated at the next order}
First we note that the terms generated at order $\alpha'^2$ quadratic in $\Delta$ will not be needed since they are of fourth power in the KK vectors and we are only considering quadratic terms. Therefore the only terms we need are the ones that are linear in $\Delta$ and come from correcting the order $\alpha'$ Lagrangian $L'_1$ in (\ref{eq:L1prime}) and (\ref{eq:L1}).\footnote{Note that we are doing the redefinition in the reduced theory, but the relevant part of the Lagrangian, i.e. the terms only involving $g,B,\Phi$, take the same form as in the unreduced theory.} To calculate these terms we use the fact that, to leading order, $H$ and the Riemann tensor change by
\begin{equation}
\delta H_{abc}=3\nabla_{[a}\Delta_{bc]}
\qquad
\delta R_{abcd}=
-\frac12\nabla_c\nabla_a\Delta_{(bd)}
+\frac14\Delta_{(ae)}R^e{}_{bcd}
-(a\leftrightarrow b)
-(c\leftrightarrow d)\,.
\end{equation}
Using these expressions we get, after a bit of work (it is easiest to start from the expression for $L_1$ in (\ref{eq:L1ord}))\footnote{Adding the same terms with the sign of $H$ changed and $F$ and $\hat F$ exchanged simplifies the expression and makes the symmetry under $B\rightarrow-B$ explicit (note that $\Delta_{ab}\rightarrow\Delta_{ba}$).}
\begin{equation}
L'_1
\rightarrow
\delta l_1
+(H\rightarrow-H,\, F\leftrightarrow\hat F)
+\mathcal O(\alpha'^3)\,,
\label{eq:delta-l1-1}
\end{equation}
with
\begin{equation}
\begin{split}
\delta l_1
&=
-\Delta_{ac}R^{adef}R^c{}_{def}%
-2\nabla_c\nabla_a\Delta_{(bd)}R^{abcd}%
-\nabla_a\Delta_{[be]}H^{ecd}R^{ab}{}_{cd}%
-\frac12\nabla_e\Delta_{ab}H^{ecd}R^{ab}{}_{cd}%
\\
&\quad
+\frac12\nabla_c\nabla_a\Delta_{(bd)}(H^2)^{abcd}%
+\frac34\Delta^{(ac)}(H^2)_{adef}R_c{}^{def}%
+\frac14\Delta^{ab}H_{acd}H_{bef}R^{cdef}
\\
&\quad
+\Delta_{ab}H^2_{cd}R^{acbd}
+\frac12(\nabla^c-2\partial^c\Phi)\nabla_c\Delta^{ab}H^2_{ab}
-\nabla^a[(\nabla_c-2\partial_c\Phi)\Delta^{(bc)}]H^2_{ab}
\\
&\quad
+\frac34\nabla_{[a}\Delta_{ef]}H_b{}^{ef}(H^2)^{ab}%
-\frac18\Delta^{ab}H^2_{ac}(H^2)_b{}^c
-\frac14\Delta^{ab}(H^2)^{cd}H^2_{acbd}%
\\
&\quad
+\frac18\Delta^{ab}H^2_{adef}(H^2)_b{}^{efd}%
+\frac14\nabla^{[a}\Delta^{be]}H_e{}^{cd}H^2_{acbd}%
+\Delta^{(ab)}H^2_{ac}\mathbbm G_b{}^c
+2\Delta^{ab}H^2_{acbd}\mathbbm G^{cd}%
\\
&\quad
-6\nabla_{[a}\Delta_{ef]}H_b{}^{ef}\mathbbm G^{ab}%
-\Delta^{ab}\mathbbm B_{ac}\mathbbm B_b{}^c%
-\Delta^{(ab)}\nabla_aH_{bcd}\mathbbm B^{cd}%
\\
&\quad
-(\nabla_a-2\partial_a\Phi)\Delta^{(ab)}H_{bcd}\mathbbm B^{cd}%
-2\nabla_b\Delta_{(ce)}H^{bed}\mathbbm B^c{}_d%
+3(\nabla^b-2\partial^b\Phi)\nabla_{[b}\Delta_{cd]}\mathbbm B^{cd}\,,%
%
%
\end{split}
\label{eq:delta-l1-2}
\end{equation}
which, after integrating by parts and using the form of $\Delta$ in (\ref{eq:Delta}), can be written
\begin{equation}
\begin{split}
\delta l_1
&=
-4F^{ab}\cdot\hat F_{bc}\mathcal R_{adef}\mathcal R^{cdef}
-4F_{ab}\cdot\hat F^{bc}\nabla_dH_{cef}R^{adef}
+2F^{ab}\cdot\hat F_{bc}\nabla_dH_{aef}\nabla^dH^{cef}
\\
&\quad
+\frac16F^{ab}\cdot\hat F_{ab}\nabla_cH_{def}\nabla^cH^{def}
+2F_{ab}\cdot\hat F^b{}_cH^2_{ef}R^{aecf}
-\frac12F^{ab}\cdot\hat F_{ab}H^2_{cdef}R^{cdef}
\\
&\quad
+\frac16F_{ab}\cdot\hat F_{cd}H^{acd}\nabla^bH^2
-2F^{ab}\cdot\hat F_{bc}H^2_{adef}(\nabla H)^{cdef}
-F_{ab}\cdot\hat F^b{}_cH^2_{ef}\nabla^eH^{fac}
\\
&\quad
-\frac16F_{ab}\cdot\hat F^b{}_cH^{acd}\nabla_dH^2
+F^{ab}\cdot\hat F_{bc}H^2_{adef}(H^2)^{cefd}
+\frac18F^{ab}\cdot\hat F_{ab}H^2_{cd}(H^2)^{cd}
\\
&\quad
+\delta l_1^{(\mathrm{e.o.m.})}\,,
%
%
%
\end{split}
\label{eq:delta-l1}
\end{equation}
where we have again dropped total derivatives and also terms of higher than second order in $\alpha'$. The terms proportional to the lowest order equations of motion will not be very important for our discussions as they can be canceled by field redefinitions at order $\alpha'^2$ (up to terms of yet higher order). For completeness they are given in (\ref{eq:delta-l1-eom}).

\section{Second order: Riemann cubed correction}\label{sec:R3}
At the second order in $\alpha'$ we expect a correction involving the Riemann tensor cubed. We have seen that we should use instead the torsionful Riemann tensor $\mathcal R_{abcd}$ defined in (\ref{eq:calR}). From its reduction in (\ref{eq:Rred}) we see that there is again only one contraction of the indices which does not violate $O(d,d)$ already at leading order in the number of fields, namely where we contract the first (last) pair of indices only with the first (last) pair on the other tensors, i.e.
\begin{equation}
\mathcal R_{abde}\mathcal R^{bce}{}_f\mathcal R_c{}^{afd}=t_6t_6\mathcal R^3\,.
\end{equation}
We can think of this as taking the trace over the first pair of indices and over the last pair of indices independently. Using (\ref{eq:Rred}) the dimensional reduction gives one $O(d,d)$ violating term quadratic in the vectors
\begin{equation}
t_6t_6\mathcal R^3
\rightarrow
3\hat F_{ab}\cdot F_{de}\mathcal R^{bce}{}_f\mathcal R_c{}^{afd}\,.
\end{equation}
As before, we can only cancel this by adding $H^2\mathcal R^2$ terms. The natural term to try is
\begin{equation}
H^2_{abde}\mathcal R^{bce}{}_f\mathcal R_c{}^{afd}=t_6t_6H^2\mathcal R^2\,,
\end{equation}
whose reduction gives
\begin{equation}
\begin{split}
t_6t_6H^2\mathcal R^2
&\rightarrow
(\hat F_{ab}\cdot F_{de}+F_{ab}\cdot \hat F_{de})\mathcal R^{bce}{}_f\mathcal R_c{}^{afd}
-2\hat F_{bg}\cdot\nabla^{(-)}_cF_{fd}H^{gd}{}_e\mathcal R^{bcef}
\\
&\quad
+2F_{eg}\cdot\nabla_f^{(+)}\hat F_{ca}H^{ga}{}_b\mathcal R^{bcef}
+2\hat F^{bc}\cdot F_{ef}H^2_{abd}{}^e\mathcal R_c{}^{afd}\,.
\end{split}
\end{equation}
The first term looks promising, but just as at order $\alpha'$ we must first deal with the second and third term (which are the same up to a change in the sign of $B$), which have derivatives sitting on the KK fields strengths $F$ and $\hat F$. Integrating these by parts gives (unlike at order $\alpha'$ using first the Bianchi identity leads to unwanted terms that cannot be canceled)
\begin{equation}
\begin{split}
&
\nabla_e\hat F_{ab}\cdot F_{cd}H^{def}\mathcal R^{abc}{}_f
+2\hat F_{ab}\cdot F_{ed}\nabla_cH^{fad}\mathcal R^{bce}{}_f
-\hat F_{ab}\cdot F_{cd}H^{acg}H^{def}\mathcal R^b{}_{efg}
\\
&
+\hat F^{ab}\cdot F^{cd}H^2_{aefc}\mathcal R_b{}^{fe}{}_d
-2\hat F_{ab}\cdot F_{cd}H^{ad}{}_e(\nabla_f-2\partial_f\Phi)\mathcal R^{bfec}
+(H\rightarrow-H,\, F\leftrightarrow\hat F)\,,
\end{split}
\end{equation}
where as usual, we dropped total derivative terms, i.e. terms of the form $(\nabla_a-2\partial_a\Phi)Y^a$ for some $Y^a$. The first of these terms can be integrated by parts again to give
\begin{equation}
\begin{split}
&
\frac12\hat F_{ab}\cdot\nabla^eF_{cd}H^{fcd}\mathcal R^{ab}{}_{ef}%
+\frac12\hat F_{ab}\cdot F_{cd}H^{cef}\nabla^d\mathcal R^{ab}{}_{ef}%
+\frac32\hat F_{ab}\cdot F^{cd}H_d{}^{ef}\nabla_{[c}\mathcal R^{ab}{}_{ef]}%
\\
&
-\hat F_{ab}\cdot F_{cd}\mathcal R^{abec}\mathbbm B^d{}_e\,.%
\end{split}
\end{equation}
Putting these results together and noting that the torsionful Riemann tensor satisfies
\begin{equation}
\nabla^{(+)}_{[d}\mathcal R^{ef}{}_{ef]}=-H_{[de}{}^g\mathcal R^{ef}{}_{f]g}\,,
\qquad
\nabla^{(-)[d}\mathcal R^{ef]}{}_{ef}=H^{[de}{}_g\mathcal R^{f]g}{}_{ef}
\label{eq:bianchi-id}
\end{equation}
and
\begin{equation}
(\nabla^{(-)}_c-2\partial_c\Phi)\mathcal R^{bc}{}_{ef}+H^b{}_{cd}\mathcal R^{cd}{}_{ef}
=
-2\nabla_{[e}(\mathbbm G^b{}_{f]}+\frac12\mathbbm B^b{}_{f]})
-H^b{}_{c[e}(\mathbbm G^c{}_{f]}+\frac12\mathbbm B^c{}_{f]})\,,
\label{eq:divR-id}
\end{equation}
we find that taking the combination
\begin{equation}
L_{2,1}=\frac13t_6t_6\mathcal R^3-\frac12t_6t_6H^2\mathcal R^2
\end{equation}
the terms with no $H$ cancel out and we get on reduction the following $O(d,d)$ violating terms
\begin{equation}
\begin{split}
L_{2,1}
&\rightarrow
-\hat F_{ab}\cdot F_{cd}(\nabla H)^{aefc}\mathcal R_S^b{}_{ef}{}^d
-\hat F_{ab}\cdot F_{cd}\nabla^eH^{fac}\mathcal R^b{}_{ef}{}^d
-\frac14\hat F_{ab}\cdot\nabla^eF_{cd}H^{fcd}\mathcal R^{ab}{}_{ef}
\\
&\quad
-\frac14\hat F_{ab}\cdot F_{cd}H^{cef}\nabla^d\mathcal R^{ab}{}_{ef}
+\frac34\hat F_{ab}\cdot F_{cd}H^{acg}H^{bef}\mathcal R_{efg}{}^d
-\frac12\hat F^{ab}\cdot F^{cd}H^2_{aefc}\mathcal R_b{}^{fe}{}_d
\\
&\quad
+\frac12\hat F_{ab}\cdot F_{cd}(H^2)^{ac}{}_{ef}\mathcal R^{befd}
+2\hat F^{ab}\cdot F^{cd}H_{ac}{}^e\nabla^{(+)}_{[e}(\mathbbm G_{d]b}-\frac12\mathbbm B_{d]b})
\\
&\quad
-\hat F^{ab}\cdot F^{cd}H^2_{acde}(\mathbbm G_b{}^e+\frac12\mathbbm B_b{}^e)
+\frac12\hat F_{ab}\cdot F_{cd}\mathcal R^{abec}\mathbbm B^d{}_e
+(H\rightarrow-H,\, F\leftrightarrow\hat F)\,.
\end{split}
\end{equation}
The next step is to try to cancel the third and fourth term, which have a different placement of derivatives compared to the other terms. Again we need terms of the form $H^2\mathcal R^2$. Consider the terms (we need two terms for the symmetry under $B\rightarrow-B$)
\begin{equation}
\begin{split}
H^2_{ab}(\mathcal R^a{}_{cde}\mathcal R^{bcde}+\mathcal R_{de}{}^a{}_c\mathcal R^{debc})
&\rightarrow
4\hat F_{ag}\cdot F^{bg}\mathcal R^{acde}\mathcal R_{bcde}%
-2F^{ef}\cdot\nabla^{(+)}_g\hat F_{cd}H_{bef}\mathcal R^{cdbg}%
\\
&\quad
+2\hat F^{ac}\cdot F^{de}H^2_{ab}\mathcal R^b{}_{cde}%
+(H\rightarrow-H,\, F\leftrightarrow\hat F)\,,
%
%
\end{split}
\end{equation}
where the second term can be integrated by parts to give
\begin{equation}
\begin{split}
&
-2\hat F_{ab}\cdot\nabla_eF_{cd}H^{fcd}\mathcal R^{ab}{}_{ef}%
+2\hat F_{ab}\cdot F_{cd}(\nabla H)^{cdef}\mathcal R^{ab}{}_{ef}%
+\hat F^{ab}\cdot F_{cd}H^2_{abef}\mathcal R^{efcd}%
\\
&
+4\hat F_{ab}\cdot F^{cd}H^{abe}\nabla^{(+)}_c(\mathbbm G_{ed}+\frac12\mathbbm B_{ed})
-2\hat F_{ab}\cdot F_{cd}H^{abe}H^{fcd}(\mathbbm G_{ef}+\frac12\mathbbm B_{ef})\,.
\end{split}
\end{equation}
The first term is of the right form to cancel one of our unwanted terms from before. To cancel the other unwanted term we consider a term
\begin{equation}
\begin{split}
H_{abc}H_{def}\mathcal R^{fgab}\mathcal R^{dec}{}_g
&\rightarrow
-2F^{ab}\cdot\nabla^{(+)}_b\hat F^{cd}H_{def}\mathcal R^{ef}{}_{ac}
-\frac12\nabla^{(-)}_fF_{ab}\cdot\nabla^{(+)}_c\hat F_{de}H^{abc}H^{def}
\\
&\quad
-\hat F_{ab}\cdot\nabla^{(-)e}F_{cd}H^{fcd}\mathcal R^{ab}{}_{ef}
+\hat F^{cd}\cdot F_{ab}H^{abg}H_{def}\mathcal R^{ef}{}_{cg}
\\
&\quad
+(H\rightarrow-H,\, F\leftrightarrow\hat F)\,.
\end{split}
\end{equation}
The third term is again of the same form (plus a term of higher order in fields). The second term can be integrated by parts twice to remove the derivatives from the $F$'s, however, this will generate a number of new terms and it turns out to be more convenient to add instead another term such that this term cancels out. For this reason we consider instead the combination of terms
\begin{equation}
\begin{split}
H_{abc}H_{def}\mathcal R^{fgab}\mathcal R^{dec}{}_g
&-2H_{abc}H_{def}(\nabla H)^{fgab}(\nabla H)^{dec}{}_g
\\
&\rightarrow
-2F^{ab}\cdot\nabla^{(+)}_b\hat F^{cd}H_{def}\mathcal R_{ac}{}^{ef}
-\hat F_{ab}\cdot\nabla^{(-)e}F_{cd}H^{fcd}\mathcal R_{ef}{}^{ab}
\\
&\quad
-\hat F_{ab}\cdot F^{cd}H^{abg}H_c{}^{ef}\mathcal R_{efdg}
+(H\rightarrow-H,\, F\leftrightarrow\hat F)\,.
\end{split}
\end{equation}
This leaves us to deal with the first term, which can be rewritten in two ways. Either (1), by integrating by parts yielding
\begin{equation}
\begin{split}
&
2F_{ab}\cdot\hat F_{cd}\nabla^aH^{cef}\mathcal R^{bd}{}_{ef}%
-F_{ab}\cdot\hat F_{cd}H^{cef}\nabla^d\mathcal R^{ab}{}_{ef}%
+3F_{ab}\cdot\hat F_{cd}H^{cef}\nabla^{[a}\mathcal R^{bd]}{}_{ef}%
\\
&
-F_{ab}\cdot\hat F_{cd}H^{acg}H^{def}\mathcal R^b{}_{gef}
+F^{ab}\cdot\hat F^{cd}H^2_{bdef}\mathcal R_{ac}{}^{ef}
+2(\nabla^b-2\partial^b\Phi)F_{ab}\cdot\hat F_{cd}H^{def}\mathcal R^{ac}{}_{ef}%
\end{split}
\end{equation}
or (2), by using the Bianchi identity and then integrating by parts yielding
\begin{equation}
\begin{split}
&
-2\hat F_{ab}\cdot\nabla_dF^{bc}H^{def}(\nabla H)^a{}_{cef}%
+F_{ab}\cdot\hat F_{cd}H^{cef}\nabla^d\mathcal R^{ab}{}_{ef}%
-\hat F_{cd}\cdot F_{ab}(\nabla H)^{cdef}\mathcal R^{ab}{}_{ef}%
\\
&
+2F_{ab}\cdot\hat F^{bc}\nabla_dH_{cef}\mathcal R^{adef}%
-F_{ab}\cdot\hat F_{cd}H^{acg}H^{def}\mathcal R^{bg}{}_{ef}
+F^{ab}\cdot\hat F^{cd}H^2_{bdef}\mathcal R_{ac}{}^{ef}
\\
&
-\frac12F^{ab}\cdot\hat F_{bc}H^{def}\nabla_dH^2_{aef}{}^c
-(\nabla^c-2\partial^c\Phi)\hat F_{cd}\cdot F_{ab}H^{def}\mathcal R^{ab}{}_{ef}%
+F^{ab}\cdot\hat F_{bc}\mathcal R_S^c{}_{aef}\mathbbm B^{ef}%
\\
&
+2F^{ab}\cdot\hat F_{bc}H^{cef}(\nabla^d-2\partial^d\Phi)\mathcal R_{adef}\,.
\end{split}
\end{equation}
Note that in both expressions the second term is the term we are trying to cancel. 

Putting these results together we find that the following combination of terms
\begin{equation}
\begin{split}
L_{2,2}
&=
L_{2,1}
-\frac12H^2_{ab}(\mathcal R^a{}_{cde}\mathcal R^{bcde}+\mathcal R_{de}{}^a{}_c\mathcal R^{debc})
+\frac34H_{abc}H_{def}\mathcal R^{fgab}\mathcal R^{dec}{}_g
\\
&\quad
-\frac32H_{abc}H_{def}(\nabla H)^{fgab}(\nabla H)^{dec}{}_g
\end{split}
\end{equation}
gives rise to the following $O(d,d)$ violating terms on reduction
\begin{equation}
\begin{split}
L_{2,2}
&\rightarrow
2F^{ab}\cdot\hat F_{bc}\mathcal R_{adef}\mathcal R^{cdef}%
-3\hat F_{ab}\cdot F_{cd}(\nabla H)^{aefc}\mathcal R^d{}_{ef}{}^b%
+\frac34\hat F_{ab}\cdot F_{cd}(\nabla H)^{abef}\mathcal R^{cd}{}_{ef}%
\\
&\quad
+\frac12\hat F_{ab}\cdot F^{bc}\nabla^dH^{aef}\mathcal R_{cdef}%
+\frac54\hat F_{ab}\cdot F_{cd}\nabla_eH_f{}^{ac}\nabla^eH^{fbd}%
\\
&\quad
-3\hat F_{ab}\cdot F_{cd}(\nabla H)^{acef}(\nabla H)^{bd}{}_{ef}%
-\frac34\hat F_{ab}\cdot F_{cd}\nabla^aH^{cef}\nabla^dH^b{}_{ef}%
\\
&\quad
+\frac52\hat F_{ab}\cdot F_{cd}(\nabla H)^{abef}(\nabla H)^{cd}{}_{ef}%
+\frac12\hat F_{ab}\cdot\nabla^aF_{cd}H^{bef}(\nabla H)^{cd}{}_{ef}%
\\
&\quad
+\frac32\hat F^{ab}\cdot\nabla_eF^{cd}H_{fcd}(\nabla H)^{ef}{}_{ab}%
-\frac12\hat F_{ab}\cdot\nabla_dF^{bc}H^{def}(\nabla H)^a{}_{cef}%
\\
&\quad
+k_{2,2}^{(H^2\mathcal R)}
+k_{2,2}^{(H^2\nabla H)}
+k_{2,2}^{(H^4)}
+k_{2,2}^{(\mathrm{e.o.m.})}
+(H\rightarrow-H,\, F\leftrightarrow\hat F)\,,
\label{eq:l22}
\end{split}
\end{equation}
where, due to the length of the expression, we are displaying only the terms of fourth order in fields. The terms of fifth and sixth order are given by
\begin{equation}
\begin{split}
k_{2,2}^{(H^2\mathcal R)}
&=
-\frac14\hat F_{ab}\cdot F_{cd}H^{abg}H^{cef}\mathcal R^d{}_{gef}%
-\frac12\hat F^{ab}\cdot F^{cd}H^2_{aefc}\mathcal R_b{}^{fe}{}_d%
-\hat F^{ab}\cdot F^{cd}H^2_{aefc}\mathcal R_{bd}{}^{ef}%
\\
&\quad
+\frac12\hat F^{ab}\cdot F^{cd}H^2_{acef}\mathcal R_{bd}{}^{ef}%
-\frac14\hat F^{ab}\cdot F^{cd}H^2_{cefd}\mathcal R^{ef}{}_{ab}%
-\frac58\hat F^{ab}\cdot F^{cd}H^2_{abef}\mathcal R_{cd}{}^{ef}%
\\
&\quad
+\hat F_{ab}\cdot F^{cd}(H^2)^{ae}\mathcal R^b{}_{ecd}%
-\frac14\hat F_{ab}\cdot F^{bc}H^{aef}H_{cgh}\mathcal R^{gh}{}_{ef}%
+\frac12\hat F^{ab}\cdot F_{bc}H^2_{adef}\mathcal R^{cefd}\,,%
\end{split}
\end{equation}
\begin{equation}
\begin{split}
k_{2,2}^{(H^2\nabla H)}
&=
-\hat F^{ab}\cdot F^{cd}H^2_{aefc}\nabla_bH_d{}^{ef}%
-\frac34\hat F^{ab}\cdot F^{cd}H_{ac}{}^e\nabla_bH^2_{de}%
+\frac38\hat F^{ab}\cdot F^{cd}H_{ace}\nabla^eH^2_{bd}%
\\
&\quad
-\frac12\hat F_{ab}\cdot F_{cd}H^{abg}H^{cef}(\nabla H)^d{}_{gef}%
-\frac34\hat F^{ab}\cdot F^{cd}H^2_{abef}(\nabla H)_{cd}{}^{ef}
\\
&\quad
-\frac14\hat F^{ab}\cdot F_{bc}H^2_{adef}(\nabla H)^{cdef}\,,%
\end{split}
\end{equation}
and
\begin{equation}
k_{2,2}^{(H^4)}
=
\frac14\hat F^{ab}\cdot F^{cd}(H^2)_{ac}{}^{ef}H^2_{befd}%
+\frac18\hat F_{ab}\cdot F_{cd}H^{eac}H^{fbd}H^2_{ef}%
\,.
\end{equation}
The terms proportional to the lowest order equations of motion are given in (\ref{eq:l22-eom}).

The first term in (\ref{eq:l22}) is the only remaining term of the form $F^2\mathcal R^2$. It cannot be canceled by adding other terms. Instead it cancels against a term coming from the field redefinition carried out in the order $\alpha'$ action, namely the first term in (\ref{eq:delta-l1}), for a suitable choice of the relative coefficient of the order $\alpha'$ and order $\alpha'^2$ correction. This shows that the Riemann cubed correction is not an independent correction as far as $O(d,d)$ is concerned, but is rather part of the correction that starts as Riemann squared. Next we have three terms of the form $F^2\nabla H\mathcal R$. Note that these terms have no derivatives on the KK field strengths. The only terms that only give rise to terms without derivatives on the KK field strengths (after integrations by parts) are terms with $\varepsilon\varepsilon$-structure. These terms are analyzed in appendix \ref{app:ee-terms} where it is shown that in order not to generate additional $F^2\mathcal R^2$ terms we should consider the combination of terms
\begin{equation}
(\varepsilon_6\varepsilon_6)'\mathcal R^3
+\frac16\left((\varepsilon_7\varepsilon_7)'H^2\mathcal R^2+24H^2\mathcal R^2\right)
-\frac13\left((\varepsilon_7\varepsilon_7)'H^2(\nabla H)^2+24H^2(\nabla H)^2\right)\,,
\end{equation}
where $H^2\mathcal R^2=H^{abc}H_{abc}\mathcal R^{de}{}_{fg}\mathcal R^{fg}{}_{de}$ and the remaining structures are defined in (\ref{eq:e6}) and (\ref{eq:e7}). The $\hat F\cdot F$ terms in the reduction are given in (\ref{eq:eeTot-red}). 

Noting also that one of the three remaining terms with $F\nabla F$-structure, the last term in (\ref{eq:l22}), can be further rewritten, using the Bianchi identity and integration by parts, as (recall again that we always drop total derivatives)
\begin{equation}
\begin{split}
F_{ab}\cdot\nabla_d\hat F^{bc}H^{def}(\nabla H)^a{}_{cef}
&=
-\hat F_{ab}\cdot\nabla^aF^{cd}H^{bef}(\nabla H)_{cdef}%
+F_{ab}\cdot\hat F_{cd}\nabla^aH^{cef}(\nabla H)^{bd}{}_{ef}%
\\
&\quad
-\hat F_{ab}\cdot F^{bc}\nabla^dH^{aef}(\nabla H)_{cdef}%
-\frac12\hat F^{ab}\cdot F_{cd}(\nabla H)_{abef}(\nabla H)^{cdef}%
\\
&\quad
+F^{ab}\cdot\hat F^{cd}H^2_{aefc}R_{bd}{}^{ef}%
-\frac12F_{ab}\cdot\hat F^{cd}H^2_{cefd}R^{abef}%
\\
&\quad
-\frac12\hat F_{ab}\cdot F^{bc}H^{aef}H_{cgh}R_{ef}{}^{gh}%
-\frac12\hat F^{ab}\cdot F_{bc}H^2_{adef}R^{cdef}%
\\
&\quad
-\frac14\hat F_{ab}\cdot F^{bc}(H^2)^{aef}{}_cH^2_{ef}%
-\frac12F_{ab}\cdot(\nabla_c-2\partial_c\Phi)\hat F^{cd}H_{def}(\nabla H)^{abef}%
\\
&\quad
+(\nabla^b-2\partial^b\Phi)F_{ab}\cdot\hat F_{cd}H^{def}(\nabla H)^{ac}{}_{ef}%
+F^{ab}\cdot\hat F_{bc}H^{cef}\nabla_e\mathbbm B_{fa}%
\\
&\quad
-F_{ab}\cdot\hat F^{bc}(H^2)^{aef}{}_c\mathbbm G_{ef}\,,%
\end{split}
\end{equation}
we find that taking the Lagrangian to be
\begin{equation}
\begin{split}
L_{2,3}
&=
L_{2,2}
-\frac{1}{64}(\varepsilon_6\varepsilon_6)'\mathcal R^3
-\frac{1}{384}\left((\varepsilon_7\varepsilon_7)'H^2\mathcal R^2+24H^2\mathcal R^2\right)
\\
&\quad
+\frac{1}{192}\left((\varepsilon_7\varepsilon_7)'H^2(\nabla H)^2+24H^2(\nabla H)^2\right)\,,
\end{split}
\end{equation}
the $F\hat F\nabla H\mathcal R$ terms cancel out and we get upon reduction
\begin{equation}
\begin{split}
L_{2,3}
&\rightarrow
-\frac12\delta l_1
+\hat F_{ab}\cdot\nabla^aF_{cd}H^{bef}(\nabla H)^{cd}{}_{ef}%
+\frac32\hat F^{ab}\cdot\nabla_eF^{cd}H_{fcd}(\nabla H)^{ef}{}_{ab}%
\\
&\quad
+\frac54\hat F_{ab}\cdot F_{cd}\nabla_eH_f{}^{ac}\nabla^eH^{fbd}%
-\frac72\hat F_{ab}\cdot F_{cd}(\nabla H)^{acef}(\nabla H)^{bd}{}_{ef}%
-\frac94\hat F_{ab}\cdot F_{cd}\nabla^aH^{cef}\nabla^dH^b{}_{ef}%
\\
&\quad
+\frac{11}{4}\hat F_{ab}\cdot F_{cd}(\nabla H)^{abef}(\nabla H)^{cd}{}_{ef}%
+\frac34\hat F_{ab}\cdot F^{cd}\nabla^eH^{fab}\nabla_eH_{fcd}%
\\
&\quad
+\frac{13}{4}\hat F_{ab}\cdot F^{bc}\nabla^dH^{aef}\nabla_dH_{cef}%
+\frac34\hat F_{ab}\cdot F^{bc}\nabla^aH^{def}\nabla_cH_{def}%
+\frac{5}{24}\hat F_{ab}\cdot F^{ab}\nabla^cH^{def}\nabla_cH_{def}%
\\
&\quad
+\mbox{Higher order terms}
+(H\rightarrow-H,\, F\leftrightarrow\hat F)\,,
\end{split}
\end{equation}
where we incorporated the terms coming from the field redefinition in the first order action (\ref{eq:delta-l1}), with the right coefficient so as to cancel the remaining $F\hat F\mathcal R^2$ term. For readability we have suppressed the terms involving five powers or more of the fields and equations of motion terms.

Now we need to cancel the remaining $F\hat F(\nabla H)^2$ terms. It turns out that this can be done by adding the following six $H^2(\nabla H)^2$ terms to the Lagrangian
\begin{equation}
\begin{aligned}
l_1&=H^2_{abcd}(\nabla H)^{acef}(\nabla H)^{bd}{}_{ef} &&&&& l_4&=H^2_{abcd}\nabla^eH^{fab}\nabla_eH_f{}^{cd}\\
l_2&=H^2_{abcd}(\nabla H)^{abef}(\nabla H)^{cd}{}_{ef} &&&&& l_5&=H^2_{ab}\nabla^aH_{def}\nabla^bH^{def}\\
l_3&=\nabla_cH^2_{ab}\nabla^c(H^2)^{ab} &&&&& l_6&=\nabla_aH^2\nabla^aH^2
\end{aligned}
\label{eq:l1-6}
\end{equation}
whose reduction is given in (\ref{eq:l1-6red}). One also has to make use of the following three identities, which are derived by integrating by parts,
\begin{equation}
\begin{split}
\hat F_{ab}\cdot F_{cd}\nabla_eH_f{}^{ac}\nabla^eH^{fbd}
&=
\hat F_{ab}\cdot F_{cd}(\nabla H)^{ac}{}_{ef}(\nabla H)^{bdef}%
-\frac12\hat F_{ab}\cdot F_{cd}(\nabla H)^{ab}{}_{ef}(\nabla H)^{cdef}%
\\
&\quad
+\frac14\hat F_{ab}\cdot F_{cd}\nabla_eH_f{}^{ab}\nabla^eH^{fcd}%
+\hat F_{ab}\cdot F_{cd}H^{acg}H^{bef}R^d{}_{gef}%
\\
&\quad
-\frac12\hat F_{ab}\cdot F_{cd}H^{abg}H^{cef}R^d{}_{gef}%
-\frac18\hat F_{ab}\cdot F_{cd}H^{eac}H^{fbd}H^2_{ef}%
\\
&\quad
+\frac{1}{16}\hat F_{ab}\cdot F_{cd}H^{eab}H^{fcd}H^2_{ef}%
+\frac12\hat F_{ab}\cdot F_{cd}\nabla_eH^{cab}\mathbbm B^{ed}%
\\
&\quad
+\frac14\hat F_{ab}\cdot F_{cd}H^{eab}\nabla_e\mathbbm B^{cd}%
-\frac12\hat F_{ab}\cdot F_{cd}H^{eac}\nabla_e\mathbbm B^{bd}%
\\
&\quad
-\frac12\hat F_{ab}\cdot F_{cd}H^{eac}H^{fbd}\mathbbm G_{ef}%
+\frac14\hat F_{ab}\cdot F_{cd}H^{eab}H^{fcd}\mathbbm G_{ef}%
\\
&\quad
+(H\rightarrow-H,\, F\leftrightarrow\hat F)\,,
\end{split}
\end{equation}
\begin{equation}
\begin{split}
\hat F_{ab}\cdot F_{cd}\nabla^aH^c{}_{ef}\nabla^dH^{bef}
&=
-\hat F_{ab}\cdot\nabla^aF^{cd}H^{bef}(\nabla H)_{cdef}%
-\frac12\hat F_{ab}\cdot F_{cd}(\nabla H)^{ab}{}_{ef}(\nabla H)^{cdef}%
\\
&\quad
-\frac14\nabla^d(\hat F_{ab}\cdot F^{bc})\nabla_d(H^2)^a{}_c%
+\frac14\nabla^c\hat F_{ab}\cdot\nabla^dF^{ab}H^2_{cd}%
\\
&\quad
+\hat F^{ab}\cdot F^{cd}H^2_{aefc}R_{bd}{}^{ef}%
-\frac14\hat F^{ab}\cdot F^{cd}H_{ace}\nabla^eH^2_{bd}%
\\
&\quad
+\frac14\hat F_{ab}\cdot F^{cd}H^{abe}\nabla_cH^2_{de}%
-\frac14\hat F^{ab}\cdot F^{cd}H^2_{ae}\nabla^eH_{bcd}%
\\
&\quad
-\frac12\hat F^{ab}\cdot F_{cd}H^2_{abef}(\nabla H)^{cdef}%
-\frac18\hat F_{ab}\cdot F^{bc}(H^2)^{ae}H^2_{ec}%
\\
&\quad
+\frac12\hat F^{ab}\cdot\nabla^c\mathbbm A_bH^2_{ac}%
-\frac12\hat F_{ab}\cdot\mathbbm A^cH_{cef}(\nabla H)^{abef}%
\\
&\quad
-\frac12\hat F_{ab}\cdot\mathbbm A^cH^{aef}\nabla^bH_{cef}%
-\frac12\hat F^{ab}\cdot F_{bc}H^2_{ad}\mathbbm G^{cd}%
\\
&\quad
-\frac14\hat F^{ab}\cdot F^{cd}H^2_{ac}\mathbbm B_{bd}%
+(H\rightarrow-H,\, F\leftrightarrow\hat F)
\end{split}
\end{equation}
and
\begin{equation}
\begin{split}
F^{ab}\cdot\hat F^{cd}\nabla_aH_c{}^{ef}\nabla_bH_{def}
&=
2\hat F_{ab}\cdot\nabla_dF^{bc}H^{aef}\nabla^dH_{cef}%
+2\hat F_{ab}\cdot\nabla^aF^{cd}H^{bef}(\nabla H)_{cdef}%
\\
&\quad
+\frac23\hat F^{ab}\cdot F_{bc}\nabla_aH_{def}\nabla^cH^{def}%
-\frac12\nabla^c\hat F^{ab}\cdot\nabla^d F_{ab}H^2_{cd}%
\\
&\quad
-\frac{1}{12}F^{ab}\cdot\nabla_c\hat F_{ab}\nabla^cH^2%
+F_{ab}\cdot\hat F^{cd}H^2_{cefd}R^{abef}%
\\
&\quad
-F^{ab}\cdot\hat F^{cd}(H^2)_c{}^eR_{abde}%
-2F_{ab}\cdot\hat F^{bc}H^2_{cdef}R^{adef}%
\\
&\quad
-F^{ab}\cdot\hat F_{bc}H^2_{ef}R_a{}^{efc}%
-\frac12\hat F_{ab}\cdot F_{cd}H^{abg}H^c{}_{ef}\nabla_gH^{def}%
\\
&\quad
+\frac12F^{ab}\cdot\nabla^e\hat F^{cd}H_{abc}H^2_{de}%
+\frac12\hat F_{ab}\cdot F^{bc}(H^2)^{ad}H^2_{cd}%
\\
&\quad
-\mathbbm A_b\cdot\nabla^d\hat F^{bc}H^2_{cd}%
-\mathbbm A^b\cdot\hat F_{cd}H^c{}_{ef}\nabla_bH^{def}%
\\
&\quad
-\frac12F^{ab}\cdot\nabla_d\hat F_{ab}H^{def}\mathbbm B_{ef}%
+2\hat F^{ab}\cdot F_{bc}H^2_{ad}\mathbbm G^{cd}%
\\
&\quad
+2\hat F^{ab}\cdot F_{bc}H_{aef}\nabla^c\mathbbm B^{ef}\,.%
\end{split}
\end{equation}
In fact, there is some redundancy in the terms in (\ref{eq:l1-6}) due to the freedom to integrate by parts and use Bianchi identities and one finds that the remaining terms of fourth order can be canceled with the following choice
\begin{equation}
sl_1-\frac{7+10s}{4}l_2+\frac{2s-3}{16}l_3+\frac{2s-1}{4}l_4+\frac{5+8s}{24}l_5+\frac{2+s}{144}l_6\,,
\end{equation}
for arbitrary $s$. We will pick $s=-\frac12$ which gives 
\begin{equation}
-\frac12l_1-\frac12l_2-\frac14l_3-\frac12l_4+\frac{1}{24}l_5+\frac{1}{96}l_6\,,
\end{equation}
because it leads to a nice form for the Lagrangian, namely
\begin{equation}
\begin{split}
L_{2,4}
&=
L_{2,3}
-\frac12H^2_{abcd}(\nabla H)^{acef}(\nabla H)^{bd}{}_{ef}
-\frac12H^2_{abcd}(\nabla H)^{abef}(\nabla H)^{cd}{}_{ef}
-\frac14\nabla_cH^2_{ab}\nabla^c(H^2)^{ab}
\\
&\quad
-\frac12H^2_{abcd}\nabla^eH^{fab}\nabla_eH_f{}^{cd}
+\frac{1}{24}H^2_{ab}\nabla^aH_{def}\nabla^bH^{def}
+\frac{1}{96}\nabla_aH^2\nabla^aH^2\,.
\end{split}
\end{equation}

After a lengthy calculation one finds that its reduction gives rise to the following $O(d,d)$ violating terms quadratic in the KK vectors
\begin{equation}
L_{2,4}
\rightarrow
-\frac12\delta l_1
+k_{2,4}^{(H^2\mathcal R)}
+k_{2,4}^{(H^2\nabla H)}
+k_{2,4}^{(H^4)}
+k_{2,4}^{(\mathrm{e.o.m.})}
+(H\rightarrow-H,\, F\leftrightarrow\hat F)\,,
\label{eq:l24}
\end{equation}
where
\begin{equation}
\begin{split}
k_{2,4}^{(H^2\mathcal R)}
&=
-\frac32\hat F_{ab}\cdot F_{cd}H^{acg}H^{bef}\mathcal R^d{}_{gef}%
-\frac54\hat F_{ab}\cdot F_{cd}H^{abg}H^{cef}\mathcal R^d{}_{gef}%
+3\hat F^{ab}\cdot F^{cd}H^2_{aefc}\mathcal R_b{}^{fe}{}_d%
\\
&\quad
-6\hat F^{ab}\cdot F^{cd}H^2_{aefc}\mathcal R_{bd}{}^{ef}%
-\frac34\hat F^{ab}\cdot F^{cd}H^2_{acef}\mathcal R_{bd}{}^{ef}%
+\frac34\hat F^{ab}\cdot F^{cd}H^2_{cefd}\mathcal R_{ab}{}^{ef}%
\\
&\quad
-\hat F^{ab}\cdot F^{cd}H^2_{abef}\mathcal R_{cd}{}^{ef}%
+\frac54\hat F^{ab}\cdot F^{cd}(H^2)_a{}^e\mathcal R_{becd}%
+\hat F_{ab}\cdot F^{bc}H^{aef}H_{cgh}\mathcal R_{ef}{}^{gh}%
\\
&\quad
+\frac52\hat F^{ab}\cdot F_{bc}H^2_{adef}\mathcal R^{cdef}%
-\frac32\hat F_{ab}\cdot F^{bc}H^2_{ef}\mathcal R^{aef}{}_c\,,%
\end{split}
\end{equation}
\begin{equation}
\begin{split}
k_{2,4}^{(H^2\nabla H)}
&=
\frac92\hat F^{ab}\cdot F^{cd}H^2_{aefc}\nabla^eH^f{}_{bd}%
-7\hat F^{ab}\cdot F^{cd}H^2_{aefc}\nabla_bH_d{}^{ef}%
-\frac34\hat F^{ab}\cdot F^{cd}H^2_{acef}\nabla_bH_d{}^{ef}%
\\
&\quad
-\hat F^{ab}\cdot F^{cd}H_{ac}{}^e\nabla_bH^2_{de}%
-\frac14\hat F^{ab}\cdot F^{cd}H_{ace}\nabla^eH^2_{bd}%
+\frac54\hat F_{ab}\cdot F^{cd}H^2_{cefd}(\nabla H)^{abef}%
\\
&\quad
-2\hat F^{ab}\cdot F^{cd}(H^2)_a{}^e(\nabla H)_{becd}%
-\frac12\hat F^{ab}\cdot F_{cd}H^2_{ae}\nabla_bH^{ecd}%
-\frac54\hat F_{ab}\cdot F^{cd}H^{abe}\nabla_cH^2_{de}%
\\
&\quad
-\frac12\hat F_{ab}\cdot F_{cd}H^2_{abef}(\nabla H)^{cdef}%
+\hat F^{ab}\cdot F_{cd}H_{abg}H^{cef}(\nabla H)^{dg}{}_{ef}%
+\frac18\hat F^{ab}\cdot F^{cd}H_{abc}\nabla_dH^2%
\\
&\quad
+\hat F_{ab}\cdot F^{bc}H^{aef}H_{cgh}(\nabla H)_{ef}{}^{gh}%
+\frac{19}{2}\hat F^{ab}\cdot F_{bc}H^2_{adef}(\nabla H)^{cdef}%
-\hat F_{ab}\cdot F^{bc}H^2_{ef}\nabla^eH^{fa}{}_c%
\\
&\quad
+\frac12\hat F_{ab}\cdot F^{bc}H^{ade}\nabla_dH^2_{ce}%
-\frac{1}{24}\hat F^{ab}\cdot F_{bc}H_a{}^{ce}\nabla_eH^2%
\end{split}
\end{equation}
and
\begin{equation}
\begin{split}
k_{2,4}^{(H^4)}
&=
\frac{11}{4}\hat F^{ab}\cdot F_{cd}H^2_{aefb}(H^2)^{cefd}%
+\frac{11}{4}\hat F^{ab}\cdot F^{cd}(H^2)_{ac}{}^{ef}H^2_{befd}%
+2\hat F^{ab}\cdot F^{cd}(H^2)_{ab}{}^{ef}H^2_{cefd}%
\\
&\quad
+\frac{3}{16}\hat F^{ab}\cdot F^{cd}H_{eac}H_{fbd}H^2_{ef}%
-\frac12\hat F^{ab}\cdot F^{cd}H^2_{acde}(H^2)_b{}^e%
+\frac34\hat F^{ab}\cdot F^{cd}H^2_{abce}(H^2)_d{}^e%
\\
&\quad
+\frac34\hat F_{ab}\cdot F_{cd}H^{eab}H^{fcd}H^2_{ef}%
-\frac{5}{32}\hat F^{ab}\cdot F^{cd}H^2_{ac}H^2_{bd}%
-\frac78\hat F_{ab}\cdot F^{bc}(H^2)^{adef}H^2_{cefd}%
\\
&\quad
+\frac18\hat F_{ab}\cdot F^{bc}(H^2)^{aef}{}_cH^2_{ef}%
-\frac34\hat F_{ab}\cdot F^{bc}(H^2)^{ad}H^2_{cd}\,.%
\end{split}
\end{equation}
Finally, the terms proportional to the lowest order equations of motion are given in (\ref{eq:l24-eom}).

To cancel the $F^2H^2\mathcal R$-terms we consider the following five terms to add to the Lagrangian (note that $H^2_{ab}{}^{cd}(H^2)^{ae}{}_{fc}R^{bf}{}_{ed}=0$ and $H^2_{aefc}(H^2)^{befd}=(H^2)_{aef}{}^b(H^2)_c{}^{efd}$)
\begin{equation}
\begin{aligned}
l_7&=H^2_{abcd}(H^2)^{aefd}\mathcal R^b{}_{ef}{}^c &&&&& l_{10}&=H^2_{abcd}(H^2)^{abef}\mathcal R_{ef}{}^{cd}\\
l_8&=H^2_{abcd}(H^2)^{caef}\mathcal R_S^{bd}{}_{ef} &&&&& l_{11}&=H^2_{ab}H^2_{cd}\mathcal R^{acbd}\\
l_9&=H^2_{abcd}H^{abg}H^{cef}\mathcal R_S^d{}_{gef} &&&&&&
\end{aligned}
\label{eq:l7-11}
\end{equation}
whose reduction is given in (\ref{eq:l7-11red}). With a bit of work one finds that for the choice
\begin{equation}
\frac32l_7-\frac34l_8+\frac54l_9+\frac12l_{10}-\frac38l_{11}\,,
\end{equation}
the remaining $F^2H^2\mathcal R$-terms and $F^2H^2\nabla H$-terms cancel out. It remains only to cancel the $F^2H^4$-terms. This can be done by adding the following four terms to the Lagrangian
\begin{equation}
\begin{aligned}
l_{12}&=(H^2)_{ab}{}^{cd}(H^2)^{aefb}H^2_{cefd} &&&&& l_{14}&=H^2_{ab}H^2_{cd}(H^2)^{acbd}\\
l_{13}&=H^2_{ab}(H^2)^{adef}(H^2)^b{}_{edf} &&&&& l_{15}&=H^2_{ab}(H^2)^{bc}(H^2)_c{}^a
\end{aligned}
\label{eq:l12-15}
\end{equation}
whose reduction is given in (\ref{eq:l12-15red}). One finds that for the choice
\begin{equation}
\frac{1}{12}l_{12}+\frac{7}{16}l_{13}+\frac{1}{32}l_{14}-\frac18l_{15}\,,
\end{equation}
also the $F^2H^4$-terms cancel and we are done.

To summarize, we have found that the Lagrangian
\begin{equation}
\begin{split}
L_{2,5}
&=
\frac13\mathcal R_a{}^{bd}{}_e\mathcal R_b{}^{ce}{}_f\mathcal R_c{}^{af}{}_d
-\frac12H^2_{abde}\mathcal R^{bce}{}_f\mathcal R_c{}^{afd}
-\frac12H^2_{ab}\left(\mathcal R^a{}_{cde}\mathcal R^{bcde}+\mathcal R_{de}{}^a{}_c\mathcal R^{debc}\right)
\\
&\quad
-\frac{1}{64}(\varepsilon_6\varepsilon_6)'\mathcal R^3
-\frac{1}{384}\left((\varepsilon_7\varepsilon_7)'H^2\mathcal R^2+24H^2\mathcal R^2\right)
+\frac{1}{192}\left((\varepsilon_7\varepsilon_7)'H^2(\nabla H)^2+24H^2(\nabla H)^2\right)
\\
&\quad
+\frac34H^{abc}H_{def}\mathcal R^{dg}{}_{ab}\mathcal R^{ef}{}_{cg}
-\frac32H^{abc}H_{def}(\nabla H)^{dg}{}_{ab}(\nabla H)^{ef}{}_{cg}
\\
&\quad
-\frac12H^2_{abcd}(\nabla H)^{acef}(\nabla H)^{bd}{}_{ef}
-\frac12H^2_{abcd}(\nabla H)^{abef}(\nabla H)^{cd}{}_{ef}
-\frac14\nabla_cH^2_{ab}\nabla^c(H^2)^{ab}
\\
&\quad
-\frac12H^2_{abcd}\nabla^eH^{fab}\nabla_eH_f{}^{cd}
+\frac{1}{24}H^2_{ab}\nabla^aH_{def}\nabla^bH^{def}
+\frac{1}{96}\nabla_aH^2\nabla^aH^2
\\
&\quad
-\frac32(H^2)_{abcd}(H^2)^{bed}{}_f\mathcal R_e{}^{afc}
+\frac34H^2_{abcd}(H^2)^{acef}\mathcal R_S^{bd}{}_{ef}
+\frac54H^2_{ab}(H^2)^{adef}\mathcal R_S^b{}_{def}
\\
&\quad
+\frac12H^2_{ab}H^{acd}H^{bef}\mathcal R_{cdef}
+\frac38H^2_{ab}H^2_{cd}\mathcal R^{acbd}
+\frac{1}{12}(H^2)_{ab}{}^{cd}(H^2)^{aefb}H^2_{cefd}
\\
&\quad
-\frac{7}{16}H^2_{ab}(H^2)^{adef}(H^2)^b{}_{efd}
-\frac{1}{32}H^2_{ab}H^2_{cd}(H^2)^{acbd}
-\frac18H^2_{ab}(H^2)^{bc}(H^2)_c{}^a\,,
\end{split}
\label{eq:L25}
\end{equation}
gives only $O(d,d)$ violating terms that (up to total derivatives) either cancel against terms from the field redefinitions in the order $\alpha'$ action or are proportional to the lowest order equations of motion and can therefore be remove by field redefinitions,
\begin{equation}
L_{2,5}\rightarrow
-\frac12\delta l_1
+k_{2,5}^{(\mathrm{e.o.m.})}
+(H\rightarrow-H,\, F\leftrightarrow\hat F)\,.
\end{equation}
Including the contribution from the order $\alpha'$ Lagrangian (\ref{eq:delta-l1-1}), we have
\begin{equation}
L'_1+2L_{2,5}
\rightarrow
2k_{2,5}^{(\mathrm{e.o.m.})}
+(H\rightarrow-H,\, F\leftrightarrow\hat F)\,.
\label{eq:l25}
\end{equation}
The expression for the equations of motion terms on the RHS is quite long, it can be found in (\ref{eq:l25-eom}). From this expression one can directly read off the field redefinitions required in the reduced theory for $O(d,d)$ invariance. It should be possible to compare this to the DFT calculation in \cite{Hronek:2022dyr}, but we will not do so here. Instead we will try to simplify the Lagrangian we have found as much as we can.

\section{Simplifying the second order Lagrangian}\label{sec:simplify}
Using the expression for $L'_1$ in (\ref{eq:L1prime}) the Lagrangian we have found up to second order in $\alpha'$ takes the form
\begin{equation}
L_1-H^2_{ab}\mathbbm G^{ab}+2L_{2,5}\,,
\end{equation}
with $L_1$ given in (\ref{eq:L1}) and $L_{2,5}$ in (\ref{eq:L25}). The latter is not particularly simple, but luckily it can be simplified. First of all it is natural to cancel the second term, which is of order $\alpha'$ but proportional to the equations of motion, by a field redefinition (remember that this term was introduced only to simplify the intermediate calculations). The following redefinition of the metric and dilaton
\begin{equation}
\Delta g_{mn}=-H^2_{mn}\,,\qquad
\Delta\Phi=\frac14H^2-\frac18H^2_{mn}(H^2)^{mn}\,,
\label{eq:g-redef}
\end{equation}
does the job since the corresponding change in the lowest order Lagrangian is
\begin{equation}
\Delta L_0=H^2_{ab}\mathbbm G^{ab}+\mathcal O(\alpha'^2)\,,
\end{equation}
canceling the term in question. However, this field redefinition will also affect higher orders in $\alpha'$. We need to calculate the terms at order $\alpha'^2$. To do this we first need the second order correction to the Riemann tensor from a variation of the metric
\begin{equation}
\Delta g_{mn}=\Delta_{mn}\,.
\end{equation}
It is given by
\begin{equation}
(\Delta R_{klmn})^{(2)}=
-\frac14(\nabla_m\Delta_{pk}+\nabla_k\Delta_{pm}-\nabla_p\Delta_{mk})(\nabla_n\Delta_l{}^p+\nabla_l\Delta_n{}^p-\nabla^p\Delta_{nl})
-(k\leftrightarrow l)\,.
%
%
\end{equation}
Using this we find the second order correction to the Ricci scalar to be, after integrations by parts,
\begin{equation}
\begin{split}
(\Delta R)^{(2)}
&=
\Delta_{kl}\nabla^2\Delta^{kl}%
+\frac34\nabla_m\Delta_{kl}\nabla^m\Delta^{kl}%
+\frac12\Delta_{km}\Delta_{nl}R^{klmn}
-\frac14\nabla_m\Delta^k{}_k\nabla^m\Delta^l{}_l%
-4\partial^k\Phi\partial^l\Phi\Delta^2_{kl}%
\\
&\quad
+2\partial^k\Phi\Delta_{kl}\nabla^l\Delta^m{}_m%
+\frac12(R^{kl}+2\nabla^k\partial^l\Phi)\Delta^2_{kl}%
+\frac12(\nabla^k-2\partial^k\Phi)\Delta_{km}(\nabla_l-2\partial_l\Phi)\Delta^{lm}\,.%
\end{split}
\end{equation}
It is now straightforward to compute the variation of the lowest order Lagrangian under (\ref{eq:g-redef}). One finds
\begin{equation}
\begin{split}
\Delta L_0
&=
H^2_{ab}\mathbbm G^{ab}
-\frac14\nabla_cH^2_{ab}\nabla^c(H^2)^{ab}
+\frac{1}{72}\nabla_aH^2\nabla^aH^2
+\frac12H^2_{ab}H^2_{cd}R^{acbd}
-\frac14H^2_{ab}H^2_{cd}(H^2)^{acbd}
\\
&\quad
-\frac18H^2_{ab}(H^2)^{bc}(H^2)_c{}^a
+\frac12H^2_{ac}(H^2)^c{}_b\mathbbm G^{ab}
+\frac16\nabla_aH^2H^{abc}\mathbbm B_{bc}
+\frac12H^2_{abcd}\mathbbm B^{ab}\mathbbm B^{cd}
+\mathcal O(\alpha'^3)\,.
\end{split}
\end{equation}
Finally we need the first order variation of the order $\alpha'$ Lagrangian, which can be read off from (\ref{eq:delta-l1-1}) and (\ref{eq:delta-l1-2}) by taking $\Delta_{ab}=-H^2_{ab}$,
\begin{equation}
\begin{split}
\Delta L'_1
&=
H^2_{ab}\mathcal R^{aecd}\mathcal R^b{}_{ecd}
+H^2_{ab}\mathcal R^{cda}{}_e\mathcal R_{cd}{}^{be}
-\frac12H^2_{ab}\nabla^eH^{acd}\nabla_eH^b{}_{cd}%
\\
&\quad
-\frac16H^2_{ab}\nabla^aH^{def}\nabla^bH_{def}%
-H^{abe}(\nabla H)_{ab}{}^{cd}\nabla_cH^2_{de}%
+\nabla_cH^2_{ab}\nabla^a(H^2)^{bc}%
\\
&\quad
-\frac{1}{18}\nabla_aH^2\nabla^aH^2%
-\frac12(H^2)^{ab}H^2_{adef}R_b{}^{def}%
-\frac12H^2_{ab}H^{acd}H^{bef}R_{cdef}%
-2H^2_{ab}H^2_{cd}R^{acbd}%
\\
&\quad
-\frac12H^2_{ab}(H^2)^{adef}(H^2)^b{}_{efd}%
+\frac14(H^2)^{ab}(H^2)^{cd}H^2_{acbd}%
+\frac14(H^2)^{ab}H^2_{bc}(H^2)^c{}_a%
\\
&\quad
+8(\nabla^a-2\partial^a\Phi)\nabla_{[a}H^2_{b]c}\mathbbm G^{bc}%
-2H^2_{ab}(H^2)^{bc}\mathbbm G_c{}^a%
-4(H^2)^{ab}H^2_{acbd}\mathbbm G^{cd}%
-\frac13\partial_aH^2H^{acd}\mathbbm B_{cd}%
\\
&{}
+5\nabla_aH^2_{bc}H^{ab}{}_d\mathbbm B^{cd}%
+2H^2_{ab}\nabla^aH^{bcd}\mathbbm B_{cd}%
+2H^2_{ab}\mathbbm B^{ac}\mathbbm B^b{}_c%
+\mathcal O(\alpha'^3)\,.
\end{split}
\end{equation}

Taking these contributions into account the Lagrangian becomes, up to second order in $\alpha'$
\begin{equation}
L_0+L_1+L_{2,6}\,,
\end{equation}
where, after removing the terms proportional to equations of motion by field redefinitions, we have
\begin{equation}
\begin{split}
L_{2,6}
&=
\frac23\mathcal R_a{}^{bd}{}_e\mathcal R_b{}^{ce}{}_f\mathcal R_c{}^{af}{}_d
-H^2_{abde}\mathcal R^{bce}{}_f\mathcal R_c{}^{afd}
-\frac{1}{32}(\varepsilon_6\varepsilon_6)'\mathcal R^3
\\
&\quad
-\frac{1}{192}\left((\varepsilon_7\varepsilon_7)'H^2\mathcal R^2+24H^2\mathcal R^2\right)
+\frac{1}{96}\left((\varepsilon_7\varepsilon_7)'H^2(\nabla H)^2+24H^2(\nabla H)^2\right)
\\
&\quad
+\frac32H^{abc}H_{def}\mathcal R^{dg}{}_{ab}\mathcal R^{ef}{}_{cg}
-3H^{abc}H_{def}(\nabla H)^{dg}{}_{ab}(\nabla H)^{ef}{}_{cg}
\\
&\quad
-H^2_{abcd}(\nabla H)^{acef}(\nabla H)^{bd}{}_{ef}
-H^2_{abcd}(\nabla H)^{abef}(\nabla H)^{cd}{}_{ef}
-H^2_{abcd}\nabla^eH^{fab}\nabla_eH_f{}^{cd}
\\
&\quad
-\frac34\nabla_cH^2_{ab}\nabla^c(H^2)^{ab}
+\nabla_cH^2_{ab}\nabla^a(H^2)^{bc}
-H^{abe}(\nabla H)_{ab}{}^{cd}\nabla_cH^2_{de}
\\
&\quad
-\frac12H^2_{ab}\nabla^eH^{acd}\nabla_eH^b{}_{cd}
-\frac{1}{12}H^2_{ab}\nabla^aH_{def}\nabla^bH^{def}
-\frac{1}{48}\nabla_aH^2\nabla^aH^2
\\
&\quad
-3H^2_{abcd}(H^2)^{bed}{}_f\mathcal R_e{}^{afc}
+\frac32H^2_{abcd}(H^2)^{acef}\mathcal R_S^{bd}{}_{ef}
+2H^2_{ab}(H^2)^{adef}\mathcal R_S^b{}_{def}
\\
&\quad
+\frac12H^2_{ab}H^{acd}H^{bef}\mathcal R_{cdef}
-\frac34H^2_{ab}H^2_{cd}\mathcal R^{acbd}
+\frac16(H^2)_{ab}{}^{cd}(H^2)^{aefb}H^2_{cefd}
\\
&\quad
-\frac78H^2_{ab}(H^2)^{adef}(H^2)^b{}_{efd}
-\frac{7}{16}H^2_{ab}H^2_{cd}(H^2)^{acbd}
-\frac18H^2_{ab}(H^2)^{bc}(H^2)_c{}^a\,.
\end{split}
\label{eq:L26}
\end{equation}
This expression is still not very simple and we would like to simplify it as much as possible. It turns out that we can get rid of most of the terms involving the fifth and sixth power of the fields. To see this we note that by integrating by parts we have the following relations
\begin{equation}
\begin{split}
(H^2)_a{}^{bd}{}_e(\nabla H)_b{}^{ce}{}_f(\nabla H)_c{}^{af}{}_d
&=
\frac14H^2_{abcd}\nabla^eH^{fac}\nabla_eH_f{}^{bd}%
-\frac14H^{abe}(\nabla H)_{ab}{}^{cd}\nabla_cH^2_{de}%
\\
&\quad
-\frac34H^2_{abcd}(\nabla H)^{abef}(\nabla H)^{cd}{}_{ef}%
-\frac12H^2_{abcd}(H^2)^{bed}{}_fR_e{}^{afc}%
\\
&\quad
+\frac18H^2_{ab}(H^2)^{adef}R^b{}_{def}%
-\frac14H^2_{adef}\nabla^dH^{bef}\mathbbm B^a{}_b%
\\
&\quad
+\frac14H_{abe}H_{fcd}(\nabla H)^{abcd}\mathbbm B^{ef}\,,%
\end{split}
\end{equation}
\begin{equation}
\begin{split}
H^2_{abcd}\nabla^eH^{fac}\nabla_eH_f{}^{bd}
&=
H^2_{abcd}(H^2)^{acef}R^{bd}{}_{ef}
+\frac14H^2_{ab}(H^2)^{adef}(H^2)^b{}_{efd}
\\
&\quad
-H^2_{abcd}H^{ace}\nabla_e\mathbbm B^{bd}
-H^2_{abcd}H^{eac}H^{fbd}\mathbbm G_{ef}\,,
\end{split}
\end{equation}
\begin{equation}
\begin{split}
H^{abe}(\nabla H)_{ab}{}^{cd}\nabla_cH^2_{de}
&=
-\frac14\nabla_cH^2_{ab}\nabla^c(H^2)^{ab}
-\frac16H^2_{ab}\nabla^aH^{def}\nabla^bH_{def}
+\frac12H^2_{ab}(H^2)^{adef}R^b{}_{def}
\\
&\quad
-\frac18H^2_{ab}(H^2)^{bc}(H^2)_c{}^a
-\frac12H^2_{ab}(H^2)^{bc}\mathbbm G_c{}^a
-\frac12H^2_{ab}H^{acd}\nabla^b\mathbbm B_{cd}\,,
\end{split}
\end{equation}
\begin{equation}
\begin{split}
\nabla_cH^2_{ab}\nabla^a(H^2)^{bc}
&=
\frac{1}{36}\nabla_aH^2\nabla^aH^2%
+H^2_{ab}H^2_{cd}R^{acbd}%
-\frac14H^2_{ab}(H^2)^{bc}(H^2)_c{}^a%
\\
&\quad
-H^2_{ab}(H^2)^{bc}\mathbbm G_c{}^a
+\frac16H^{acd}\nabla_aH^2\mathbbm B_{cd}%
-H^2_{ab}\nabla^a(H^{bcd}\mathbbm B_{cd})%
\end{split}
\end{equation}
and
\begin{equation}
\begin{split}
H^2_{ab}\nabla^eH^{acd}\nabla_eH^b{}_{cd}
&=
-\frac12\nabla_cH^2_{ab}\nabla^c(H^2)^{ab}
+2H^2_{ab}(H^2)^{adef}R^b{}_{def}
+H^2_{ab}H^{acd}H^{bef}R_{cdef}
\\
&\quad
-\frac12H^2_{ab}H^2_{cd}(H^2)^{acbd}
-\frac14H^2_{ab}(H^2)^{bc}(H^2)_c{}^a
-2H^2_{ab}(H^2)^{acbd}\mathbbm G_{cd}
\\
&\quad
-H^2_{ab}(H^2)^{bc}\mathbbm G_c{}^a
-H^2_{ab}H^{acd}\nabla^b\mathbbm B_{cd}
-2H^2_{ab}H^{acd}\nabla_c\mathbbm B_d{}^b\,.
\end{split}
\end{equation}
Using these relations we can remove eight of the $H^4\mathcal R$ and $H^6$ terms. Making again the appropriate field redefinitions to remove the terms proportional to the equations of motion and using the fact that
\begin{equation}
H^{abe}\nabla_gH^f{}_{ab}H_{cd[e}\nabla^gH_{f]}{}^{cd}
=
H^2_{abcd}\nabla^eH^{fab}\nabla_eH_f{}^{cd}
-\frac14\nabla_cH^2_{ab}\nabla^c(H^2)^{ab}
\end{equation}
we find that the order $\alpha'^2$ Lagrangian now takes the form
\begin{equation}
\begin{split}
L_{2,7}
&=
\frac23\mathcal R_{abde}\mathcal R^{bce}{}_f\mathcal R_c{}^{afd}
-H^2_{abde}\mathcal R^{bce}{}_f\mathcal R_c{}^{afd}
+2(H^2)_{abde}(\nabla H)^{bce}{}_f(\nabla H)_c{}^{afd}
-\frac{1}{32}(\varepsilon_6\varepsilon_6)'\mathcal R^3
\\
&\quad
-\frac{1}{192}\left((\varepsilon_7\varepsilon_7)'H^2\mathcal R^2+24H^2\mathcal R^2\right)
+\frac{1}{96}\left((\varepsilon_7\varepsilon_7)'H^2(\nabla H)^2+24H^2(\nabla H)^2\right)
\\
&\quad
+\frac32H^{abc}H_{def}\mathcal R^{dg}{}_{ab}\mathcal R^{ef}{}_{cg}
-3H^{abc}H_{def}(\nabla H)^{dg}{}_{ab}(\nabla H)^{ef}{}_{cg}
-H^2_{abcd}(\nabla H)^{acef}(\nabla H)^{bd}{}_{ef}
\\
&\quad
+\frac12H^2_{abcd}(\nabla H)^{abef}(\nabla H)^{cd}{}_{ef}%
+H^2_{abcd}\nabla^eH^{fac}\nabla_eH_f{}^{bd}
-\frac12H^2_{abcd}\nabla^eH^{fab}\nabla_eH_f{}^{cd}
\\
&\quad
-\frac12H^{abe}\nabla_gH^f{}_{ab}H_{cd[e}\nabla^gH_{f]}{}^{cd}
+\frac12\nabla_cH^2_{ab}\nabla^{[a}(H^2)^{c]b}
+H^{abe}(\nabla H)_{ab}{}^{cd}\nabla_cH^2_{de}
\\
&\quad
+\frac16H^2_{ab}\nabla^aH^{def}\nabla^bH_{def}
-2H^2_{abcd}(H^2)^{bed}{}_f\mathcal R_e{}^{afc}
-\frac13(H^2)_{ab}{}^{cd}(H^2)^{aefb}H^2_{cefd}\,,
\end{split}
\label{eq:L27}
\end{equation}
which is quite a bit simpler. Finally, using the definitions in (\ref{eq:t6}) and (\ref{eq:e7hat}), in particular that
\begin{equation}
\begin{split}
(\widehat{\varepsilon_7\varepsilon_7})'H^2\mathcal R^2
&=
(\varepsilon_7\varepsilon_7)'H^2\mathcal R^2
+24H^2\mathcal R^2
-288H^{abc}H_{def}\mathcal R^{dg}{}_{ab}\mathcal R^{ef}{}_{cg}
\\
&=
288H^2_{ab}\mathcal R^{efad}\mathcal R^b{}_{def}
-\frac92(4!)^2(H^2)^{ab}{}_{cd}(\mathcal R^{[cd}{}_{[ab}\mathcal R^{ef]}{}_{ef]})'\,,
\end{split}
\end{equation}
we get the form of the $\alpha'^2$ Lagrangian given in (\ref{eq:L2}).

It is useful to write this result also in terms of the standard Riemann tensor. This is straightforward but tedious. In particular one has to integrate the $(\nabla H)^2R$-terms by parts to get
\begin{equation}
\begin{split}
&
-6(\nabla H)^{ab}{}_{ce}(\nabla H)^{cd}{}_{af}R^{ef}{}_{bd}
-2(\nabla H)^{ab}{}_{ce}(\nabla H)^{ef}{}_{bd}R_f{}^{cd}{}_a
+\frac32(\nabla H)^{ab}{}_{cd}(\nabla H)^{cd}{}_{ef}R^{ef}{}_{ab}
\\
&=
3H^2_{abcd}R^{aefc}R^b{}_{ef}{}^d
-2H^2_{abcd}R^{acef}R^{bd}{}_{ef}
+2H_{abc}H_{def}R^{cgef}R^{abd}{}_g
+H^2_{ac}R^{adef}R^c{}_{def}
\\
&\quad
-\frac34H^2_{abcd}R^{abef}R^{cd}{}_{ef}
-\frac{5}{24}H^2_{ab}\nabla^aH^{def}\nabla^bH_{def}
+\frac{3}{16}\nabla^cH^2_{ab}\nabla_c(H^2)^{ab}
-\frac{1}{72}\nabla^aH^2\nabla_aH^2
\\
&\quad
-H^2_{abcd}(H^2)^c{}_eR^{abde}
+\frac18H^2_{ab}H^{acd}H^{bef}R_{cdef}
-\frac12H^2_{ab}H^2_{cd}R^{acbd}
-\frac{1}{32}H^2_{ab}(H^2)^{bc}(H^2)_c{}^a
\\
&\quad
+\mbox{e.o.m. terms}\,.
\end{split}
\end{equation}
After some work one finds that the Lagrangian becomes, after using field redefinitions at second order in $\alpha'$ to remove the equation of motion terms,
\begin{equation}
\begin{split}
L_2
&=
-\frac43R_{abde}R^{bcef} R_c{}^a{}_f{}^d
+R_{abcd}R^{abef}R^{cd}{}_{ef}
-H^2_{abcd}R^{acef}R^{bd}{}_{ef}
-\frac12H^2_{abcd}R^{abef}R^{cd}{}_{ef}
\\
&\quad
-\frac12H^2_{ab}R^{acde}R^b{}_{cde}
+2H^{abc}H^{def}R_{abdg}R_{efc}{}^g
+H^2_{abcd}\nabla^eH^{fac}\nabla_eH_f{}^{bd}
\\
&\quad
-\frac12H^2_{abcd}(\nabla H)^{abef}(\nabla H)_{cd}{}^{ef}
+\frac14H^{abc}H^{def}\nabla_dH_{gab}\nabla_cH^g{}_{ef}
-\frac{1}{12}H^3_{abc}(H^3)^{abc}
\\
&\quad
-\frac14H^2_{ab}(H^2)^{adef}(H^2)^b{}_{efd}
+\frac{1}{48}H^2_{ab}(H^2)^{bc}(H^2)_c{}^a\,.
\end{split}
\end{equation}
This coincides with the second order Lagrangian found in \cite{Gholian:2023kjj}, in what was referred to there as the `Meissner scheme'. In that scheme the Lagrangian at first order in $\alpha'$ is more complicated, but one can show that there is a field redefinition which simplifies the first order Lagrangian without complicating the second order Lagrangian, consistent with our results.

\vspace{4cm}

\section*{Acknowledgments}
This work is supported by the grant ``Dualities and higher derivatives'' (GA23-06498S) from the Czech Science Foundation (GA\v CR).

\newpage

\appendix

\section{Reduction of \texorpdfstring{$\varepsilon$}{e}-terms}\label{app:ee-terms}
Here we compute the $O(d,d)$ violating terms quadratic in the KK vectors which arise in dimensional reduction of the terms with $\varepsilon\varepsilon$-structure defined in (\ref{eq:e6}) and (\ref{eq:e7}). Using (\ref{eq:Rred}) one finds
\begin{equation}
\frac13(\varepsilon_6\varepsilon_6)'\mathcal R^3
\rightarrow
\frac12(\varepsilon_6\varepsilon_6)'[\hat F\cdot F]\mathcal R^2
+4\cdot5!(\nabla^{(-)a}F_{[ab}\cdot\nabla^{(+)}_c\hat F^{bc}\mathcal R^{de}{}_{de]})'
+(H\rightarrow-H,\, F\leftrightarrow\hat F)\,,
\end{equation}
where we recall that a prime denotes the removal of terms with self-contractions, this includes contractions between $F$ and $\hat F$ in the first term (we put square brackets to emphasize that it is to be considered as a single factor). The second term needs to be integrated by parts twice to remove the derivatives from the KK field strengths. The anti-symmetrization in the upper and lower indices guarantees that this can be done. A tedious calculation gives
\begin{equation}
\frac13(\varepsilon_6\varepsilon_6)'\mathcal R^3
\rightarrow
k_1^{(\mathcal R^2)}
+k_1^{(\nabla H\mathcal R)}
+k_1^{((\nabla H)^2)}
+k_1^{(H^2\mathcal R)}
+k_1^{(H^2\nabla H)}
+k_1^{(H^4)}
+k_1^{(\mathrm{e.o.m.})}
+(H\rightarrow-H,\, F\leftrightarrow\hat F)\,,
\end{equation}
where
\begin{equation}
k_1^{(\mathcal R^2)}
=
-32\hat F^{ab}\cdot F_{cd}\mathcal R^{ce}{}_{af}\mathcal R^{df}{}_{be}
+8\hat F^{ab}\cdot F_{cd}\mathcal R^{cd}{}_{ef}\mathcal R^{ef}{}_{ab}
+32\hat F^{ab}\cdot F_{bc}\mathcal R^{ef}{}_{ad}\mathcal R^{cd}{}_{ef}\,,
\end{equation}
\begin{equation}
\begin{split}
k_1^{(\nabla H\mathcal R)}
&=
64\hat F^{ab}\cdot F^{cd}\nabla^eH^f{}_{ac}\mathcal R_{debf}
-32\hat F^{ab}\cdot F^{cd}\nabla_aH_{cef}\mathcal R_{bd}{}^{ef}
+32\hat F_{ab}\cdot F_{cd}(\nabla H)^{abef}\mathcal R^{cd}{}_{ef}
\\
&\quad
-64\hat F_{ab}\cdot F^{bc}\nabla^dH^{aef}\mathcal R_{cdef}\,,
\end{split}
\end{equation}
\begin{equation}
k_1^{((\nabla H)^2)}
=
32\hat F^{ab}\cdot F^{cd}\nabla_aH_{cef}\nabla_dH_b{}^{ef}
-\frac{64}{3}\hat F_{ab}\cdot F^{bc}\nabla^aH^{def}\nabla_cH_{def}
-\frac83\hat F_{ab}\cdot F^{ab}\nabla_cH_{def}\nabla^cH^{def}\,,
\end{equation}
\begin{equation}
\begin{split}
k_1^{(H^2\mathcal R)}
&=
-128\hat F_{ab}\cdot F_{cd}H^{acg}H^{bef}\mathcal R^d{}_{efg}
+64\hat F_{ab}\cdot F_{cd}H^{acg}H^{bef}\mathcal R^d{}_{gef}
\\
&\quad
+64\hat F_{ab}\cdot F_{cd}H^{cdg}H^{aef}\mathcal R_{gef}{}^b
-64\hat F^{ab}\cdot F^{cd}H^2_{aefc}\mathcal R_d{}^{fe}{}_b
-64\hat F^{ab}\cdot F^{cd}H^2_{aefc}\mathcal R_d{}^{ef}{}_b
\\
&\quad
+64\hat F^{ab}\cdot F^{cd}H^2_{acef}\mathcal R_d{}^{ef}{}_b
-32\hat F^{ab}\cdot F^{cd}H^2_{cefd}\mathcal R^{ef}{}_{ab}
+16\hat F^{ab}\cdot F^{cd}H^2_{cdef}\mathcal R^{ef}{}_{ab}
\\
&\quad
+32\hat F^{ab}\cdot F^{cd}H^2_{ae}\mathcal R_{cdb}{}^e
+48\hat F_{ab}\cdot F^{bc}H^{aef}H_{cgh}\mathcal R^{gh}{}_{ef}
+128\hat F^{ab}\cdot F_{bc}H^2_{adef}\mathcal R^{cdef}
\\
&\quad
+32\hat F_{ab}\cdot F^{bc}H^2_{ef}\mathcal R^{aef}{}_c
+8\hat F_{ab}\cdot F^{ab}H^2_{efgh}R^{efgh}\,,
\end{split}
\end{equation}
\begin{equation}
\begin{split}
k_1^{(H^2\nabla H)}
&=
32\hat F^{ab}\cdot F^{cd}H^2_{aefc}\nabla_bH_d{}^{ef}
+16\hat F^{ab}\cdot F^{cd}H^2_{acef}\nabla_bH_d{}^{ef}
-\frac83\hat F^{ab}\cdot F^{cd}H_{abc}\nabla_dH^2
\\
&\quad
-64\hat F_{ab}\cdot F^{bc}H^{aef}H_{cgh}(\nabla H)_{ef}{}^{gh}
+64\hat F^{ab}\cdot F_{bc}H^2_{adef}(\nabla H)^{cdef}
\\
&\quad
+32\hat F_{ab}\cdot F^{bc}H^{aef}\nabla_eH^2_{fc}
+32\hat F_{ab}\cdot F^{bc}H^2_{ef}\nabla^eH^{fa}{}_c
+\frac83\hat F_{ab}\cdot F^{bc}H^a{}_{ce}\nabla^eH^2\,,
\end{split}
\end{equation}
\begin{equation}
\begin{split}
k_1^{(H^4)}
&=
-96\hat F_{ab}\cdot F^{bc}(H^2)^{adef}H^2_{cefd}
+32\hat F_{ab}\cdot F^{bc}(H^2)^{aef}{}_cH^2_{ef}
-16\hat F_{ab}\cdot F^{bc}(H^2)^{ad}H^2_{cd}
\\
&\quad
-2\hat F_{ab}\cdot F^{ab}(H^2)^{cd}H^2_{cd}
\end{split}
\end{equation}
and
\begin{equation}
\begin{split}
k_1^{(\mathrm{e.o.m.})}
&=
-\frac{16}{3}\hat{\mathbbm A}_b\cdot F^{bc}\nabla_cH^2
-64\hat{\mathbbm A}^a\cdot F^{cd}H_c{}^{ef}\mathcal R_{defa}
+16\hat{\mathbbm A}_a\cdot F^{cd}H^{aef}\mathcal R_{cdef}
\\
&\quad
-32\nabla^a\hat{\mathbbm A}^b\cdot F^{cd}\mathcal R_{cdab}
-32\nabla^{(+)}_e\hat F^{ab}\cdot\mathbbm A_d\mathcal R^{de}{}_{ab}
-32\hat F^{ab}\cdot F^{cd}\mathcal R_{cdeb}\mathbbm G_a{}^e
\\
&\quad
-8\hat F_{ab}\cdot F^{ab}H^2_{cd}\mathbbm G^{cd}
-64\hat F_{ab}\cdot F^{bc}\mathcal R_c{}^{efa}\mathbbm B_{ef}
-8\hat F_{ab}\cdot F^{ab}H^{cef}\nabla_c\mathbbm B_{ef}
\\
&\quad
-80\hat F_{ab}\cdot F^{cd}\mathcal R^{cdbe}\mathbbm B^a{}_e
+64\hat F_{ab}\cdot F^{bc}H_{cef}[(\nabla^{(-)}_g-2\partial_g\Phi)\mathcal R^{egaf}+H^{egh}\mathcal R_{gh}{}^{af}]%
\\
&\quad
-32\hat F^{ab}\cdot F_{bc}H_{aef}[(\nabla^{(-)}_d-2\partial_d\Phi)\mathcal R^{cdef}+H^{cgh}\mathcal R_{gh}{}^{ef}]
\\
&\quad
+64\hat F_{ab}\cdot F_{cd}H^{ac}{}_f[(\nabla^{(-)}_e-2\partial_e\Phi)\mathcal R^{debf}+H^{dgh}\mathcal R_{gh}{}^{bf}]
\\
&\quad
-16F^{ab}\cdot\hat F_{cd}H_{fab}[(\nabla^{(-)}_e-2\partial_e\Phi)\mathcal R^{fecd}+H^{fgh}\mathcal R_{gh}{}^{cd}]
\\
&\quad
-64\hat F_{bd}\cdot\nabla^{(-)}_eF^{bc}[(\nabla^{(-)a}-2\partial^a\Phi)\mathcal R_{ca}{}^{de}+H_{cgh}\mathcal R^{ghde}]\,,
\end{split}
\end{equation}
where the last five terms can be written in terms of the equations of motion via (\ref{eq:divR-id}). In the calculation we made use of the identities in (\ref{eq:bianchi-id}), the fact that
\begin{equation}
\begin{split}
\nabla^{(-)[b}\nabla^{(+)}_e\hat F^{cd]}&=-\mathcal R^{[bc}{}_{eg}\hat F^{d]g}+H^{[bc}{}_g\nabla^{(+)}_e\hat F^{d]g}\,,\\
\nabla^{(+)}_{[b}\nabla^{(-)e}F_{cd]}&=-\mathcal R^{eg}{}_{[bc}F_{d]g}-H_{[bc}{}^g\nabla^{(-)e}F_{d]g}
\end{split}
\end{equation}
as well as the relation
\begin{equation}
\begin{split}
F_{ab}\cdot\nabla^b\hat F_{cd}H^{agh}\mathcal R^{cd}{}_{gh}
&=
F_{ab}\cdot\hat F_{cd}H^{agh}\nabla^b\mathcal R^{cd}{}_{gh}
-2\nabla_aF_c{}^b\cdot\hat F_{bd}H^{agh}\mathcal R^{cd}{}_{gh}
\\
&\quad
+2F_{ac}\cdot(\nabla^b-2\partial^b\Phi)\hat F_{bd}H^{agh}\mathcal R^{cd}{}_{gh}
+2F_{ac}\cdot\hat F_{bd}\nabla^bH^{agh}\mathcal R^{cd}{}_{gh}
\\
&\quad
+3F_{ac}\cdot\hat F_{bd}H^{agh}\nabla^{[b}\mathcal R^{cd]}{}_{gh}
-2F^{ab}\cdot\hat F_{bd}\nabla_cH_a{}^{gh}\mathcal R^{cd}{}_{gh}
\\
&\quad
-2F^{ab}\cdot\hat F_{bd}H_a{}^{gh}(\nabla_c-2\partial_c\Phi)\mathcal R^{cd}{}_{gh}\,,
\end{split}
\end{equation}
which is proven by using the Bianchi identity and integrating by parts.

Notice that $(\varepsilon_6\varepsilon_6)'\mathcal R^3$ gives rise to terms of the form $\hat FF\mathcal R^2$, but all terms of this structure are already accounted for, so these terms need to be canceled. This can be done by adding the $H^2\mathcal R^2$ term
\begin{equation}
\begin{split}
\frac19(\varepsilon_7\varepsilon_7)'HH\mathcal R^2
&\rightarrow
\frac12\varepsilon_6\varepsilon_6(\hat F\cdot F+F\cdot\hat F)(\mathcal R^2)'
+\frac19(\varepsilon_7\varepsilon_7)'[\hat F\cdot F]HH\mathcal R
\\
&\quad
-\frac496!H^{abc}H_{[abc}(\nabla^{(-)e}F_{fe}\cdot\nabla^{(+)}_{g]}\hat F^{fg})'
\\
&\quad
-\frac436!H_{[dbc}F^{bc}\cdot(\nabla_e^{(+)}\hat F^{de}\mathcal R^{fg}{}_{fg]})'
+(H\rightarrow-H,\, F\leftrightarrow\hat F)\,.
\end{split}
\end{equation}
Again the primes are a reminder that terms with self-contractions are to be removed (with $\hat F\cdot F$ in the second term considered as a single object). The last two terms contain derivatives of the KK field strengths, so they need to be integrated by parts. The first of these in particular produces a lot of term as it needs to be integrated by parts twice. However, we can avoid this by adding another term to the effective action which gives a contribution canceling this term. The following combination does the trick
\begin{equation}
\begin{split}
\frac19&(\varepsilon_7\varepsilon_7)'H^2\mathcal R^2
-\frac29(\varepsilon_7\varepsilon_7)'H^2(\nabla H)^2
\\
&\rightarrow
\frac12\varepsilon_6\varepsilon_6(\hat F\cdot F+F\cdot\hat F)(\mathcal R^2)'
-2\varepsilon_6\varepsilon_6\hat F\cdot F((\nabla H)^2)'
+\frac19(\varepsilon_7\varepsilon_7)'[\hat F\cdot F]H^2\mathcal R
\\
&\quad
+\frac29(\varepsilon_7\varepsilon_7)'[\hat F\cdot F]H^2\nabla H
-\frac436!H_{[dbc}F^{bc}\cdot(\nabla_e^{(+)}\hat F^{de}\mathcal R^{fg}{}_{fg]})'
\\
&\quad
-\frac836!H_{[dbc}F^{bc}\cdot(\nabla^{(+)}_e\hat F^{de}(\nabla H)^{fg}{}_{fg]})'
+(H\rightarrow-H,\, F\leftrightarrow\hat F)\,.
\end{split}
\end{equation}
The RHS is now simpler to compute, and contains less terms, since we only need to integrate one derivative by parts. It turns out to be convenient to remove from the LHS the terms involving $H^2=H_{abc}H^{abc}$, which have the form $H^2\mathcal R^2=H^2\mathcal R^{ab}{}_{cd}\mathcal R^{cd}{}_{ab}$ and similarly for the other term. Another tedious calculation then gives
\begin{equation}
\begin{split}
\frac19&\left((\varepsilon_7\varepsilon_7)'H^2\mathcal R^2+24H^2\mathcal R^2\right)
-\frac29\left((\varepsilon_7\varepsilon_7)'H^2(\nabla H)^2+24H^2(\nabla H)^2\right)
\\
&\rightarrow
k_2^{(\mathcal R^2)}
+k_2^{(\nabla H\mathcal R)}
+k_2^{((\nabla H)^2)}
+k_2^{(H^2\mathcal R)}
+k_2^{(H^2\nabla H)}
+k_2^{(H^4)}
+k_2^{(\mathrm{e.o.m.})}
+(H\rightarrow-H,\, F\leftrightarrow\hat F)\,,
%
%
\end{split}
\end{equation}
where
\begin{equation}
k_2^{(\mathcal R^2)}
=
64\hat F^{ab}\cdot F_{cd}\mathcal R_{af}{}^{ce}\mathcal R^{df}{}_{be}
-16\hat F_{ab}\cdot F_{cd}\mathcal R^{ab}{}_{ef}\mathcal R^{cdef}
-64\hat F_{ab}\cdot F^{bc}\mathcal R^{adef}\mathcal R_{cdef}\,,
\end{equation}
\begin{equation}
\begin{split}
k_2^{(\nabla H\mathcal R)}
&=
-128\hat F^{ab}\cdot F^{cd}\nabla^eH^f{}_{ac}\mathcal R_{debf}
+64\hat F^{ab}\cdot F^{cd}\nabla_aH_{cef}\mathcal R^{ef}{}_{bd}
\\
&\quad
-64\hat F^{ab}\cdot F^{cd}(\nabla H)_{abef}\mathcal R_{cd}{}^{ef}
+128\hat F^{ab}\cdot F_{bc}\nabla_dH_{aef}\mathcal R^{efcd}\,%
\end{split}
\end{equation}
\begin{equation}
\begin{split}
k_2^{((\nabla H)^2)}
&=
-128\hat F^{ab}\cdot F^{cd}(\nabla H)_{acef}(\nabla H)_{bd}{}^{ef}%
-32\hat F_{ab}\cdot F^{cd}\nabla^eH^{fab}\nabla_eH_{fcd}
\\
&\quad
+64\hat F_{ab}\cdot F^{bc}\nabla^dH^{aef}\nabla_dH_{cef}%
-\frac{64}{3}\hat F_{ab}\cdot F^{bc}\nabla^aH^{def}\nabla_cH_{def}\,,%
\end{split}
\end{equation}
\begin{equation}
\begin{split}
k_2^{(H^2\mathcal R)}
&=
-64\hat F_{ab}\cdot F_{cd}H^{acg}H^{bef}\mathcal R^d{}_{gef}
+64\hat F^{ab}\cdot F^{cd}H^2_{aefc}\mathcal R_b{}^{fe}{}_d
\\
&\quad
-16\hat F^{ab}\cdot F^{cd}H^2_{abef}\mathcal R^{ef}{}_{cd}
-32\hat F^{ab}\cdot F^{cd}(H^2)_a{}^e\mathcal R_{becd}\,,
\end{split}
\end{equation}
\begin{equation}
\begin{split}
k_2^{(H^2\nabla H)}
&=
64\hat F^{ab}\cdot F^{cd}H^2_{aefc}\nabla_bH_d{}^{ef}
+32\hat F^{ab}\cdot F^{cd}H^2_{acef}\nabla_bH_d{}^{ef}
\\
&\quad
-64\hat F_{ab}\cdot F^{cd}H^{abg}H_{cef}(\nabla H)_{dg}{}^{ef}
-128\hat F^{ab}\cdot F_{bc}H^2_{adef}(\nabla H)^{cdef}%
\\
&\quad
-64\hat F_{ab}\cdot F^{bc}H^{aef}\nabla_eH^2_{fc}\,,%
\end{split}
\end{equation}
\begin{equation}
\begin{split}
k_2^{(H^4)}
&=
64\hat F^{ab}\cdot F^{cd}H^2_{aefc}(H^2)_{bd}{}^{ef}
-32\hat F_{ab}\cdot F^{cd}(H^2)^{abef}H^2_{cefd}
\\
&\quad
-8\hat F_{ab}\cdot F_{cd}H^{abe}H^{fcd}H^2_{ef}
+16\hat F_{ab}\cdot F_{cd}H^{eac}H^{fbd}H^2_{ef}
\end{split}
\end{equation}
and
\begin{equation}
\begin{split}
k_2^{(\mathrm{e.o.m.})}
&=
64\mathbbm A_a\cdot\hat F^{cd}H_{cef}\mathcal R^{ad}{}_{ef}
-128F_{ab}\cdot\hat{\mathbbm A}_dH^{aef}\mathcal R_S^{bd}{}_{ef}%
+32\hat F^{cd}\cdot\mathbbm A^aH_{aef}\mathcal R^{ef}{}_{cd}
\\
&\quad
+256F_{ab}\cdot\hat F^{bc}H^{aef}\nabla_e\mathbbm G_{cf}%
-64F^{ab}\cdot\hat F_b{}^cH^2_{aefc}\mathbbm B^{ef}%
-64F_c{}^b\cdot\hat F_{bd}\mathcal R^{cd}{}_{ef}\mathbbm B^{ef}%
\\
&\quad
+64F_{ab}\cdot\hat F^{cd}\mathcal R^{ab}{}_{ec}\mathbbm B_d{}^e
-128F_{ab}\cdot\hat F_{cd}\mathcal R^{bdce}\mathbbm B^a{}_e
\\
&\quad
+64F^{ab}\cdot\hat F_{cd}H_{fab}[(\nabla^{(-)}_e-2\partial_e\Phi)\mathcal R^{cefd}+H^{cgh}\mathcal R_{gh}{}^{fd}]
\\
&\quad
-128F_{ab}\cdot\hat F_{cd}H_f{}^{ac}[(\nabla^{(-)}_e-2\partial_e\Phi)\mathcal R^{befd}+H^{bgh}\mathcal R_{gh}{}^{fd}]\,.
\end{split}
\end{equation}

Putting these results together and simplifying a bit we have
\begin{equation}
\begin{split}
\frac13&(\varepsilon_6\varepsilon_6)'\mathcal R^3
+\frac{1}{18}\left((\varepsilon_7\varepsilon_7)'H^2\mathcal R^2+24H^2\mathcal R^2\right)
-\frac19\left((\varepsilon_7\varepsilon_7)'H^2(\nabla H)^2+24H^2(\nabla H)^2\right)
\\
&\rightarrow
k_3^{(\nabla H\mathcal R)}
+k_3^{((\nabla H)^2)}
+k_3^{(H^2\mathcal R)}
+k_3^{(H^2\nabla H)}
+k_3^{(H^4)}
+k_3^{(\mathrm{e.o.m.})}
+(H\rightarrow-H,\, F\leftrightarrow\hat F)\,,
\end{split}
\label{eq:eeTot-red}
\end{equation}
where
\begin{equation}
k_3^{(\nabla H\mathcal R)}
=
64\hat F^{ab}\cdot F_{cd}(\nabla H)_{ae}{}^{fc}\mathcal R^{de}{}_{bf}%
+16\hat F^{ab}\cdot F^{cd}(\nabla H)_{abef}\mathcal R_{cd}{}^{ef}%
+64\hat F_{ab}\cdot F^{bc}(\nabla H)^{adef}\mathcal R_{cdef}\,,%
\end{equation}
\begin{equation}
\begin{split}
k_3^{((\nabla H)^2)}
&=
32\hat F^{ab}\cdot F^{cd}\nabla_aH_{cef}\nabla_dH_b{}^{ef}%
-16\hat F_{ab}\cdot F^{cd}\nabla^eH^{fab}\nabla_eH_{fcd}%
-32\hat F_{ab}\cdot F^{bc}\nabla^dH^{aef}\nabla_dH_{cef}%
\\
&\quad
-\frac{32}{3}\hat F_{ab}\cdot F^{bc}\nabla^aH^{def}\nabla_cH_{def}%
-\frac{8}{3}\hat F_{ab}\cdot F^{ab}\nabla^cH^{def}\nabla_cH_{def}\,,%
\end{split}
\end{equation}
\begin{equation}
\begin{split}
k_3^{(H^2\mathcal R)}
&=
96\hat F_{ab}\cdot F_{cd}H^{acg}H^{bef}\mathcal R^d{}_{gef}%
+32\hat F_{ab}\cdot F_{cd}H^{abg}H^{cef}\mathcal R^d{}_{gef}%
-96\hat F^{ab}\cdot F^{cd}H^2_{aefc}\mathcal R_b{}^{fe}{}_d%
\\
&\quad
+64\hat F^{ab}\cdot F^{cd}H^2_{aefc}\mathcal R_{bd}{}^{ef}%
+32\hat F^{ab}\cdot F^{cd}H^2_{acef}\mathcal R_{bd}{}^{ef}%
-32\hat F^{ab}\cdot F^{cd}H^2_{cefd}\mathcal R^{ef}{}_{ab}%
\\
&\quad
+8\hat F^{ab}\cdot F^{cd}H^2_{abef}\mathcal R_{cd}{}^{ef}%
+16\hat F^{ab}\cdot F^{cd}(H^2)_a{}^e\mathcal R_{becd}%
+48\hat F_{ab}\cdot F^{bc}H^{aef}H_{cgh}\mathcal R^{gh}{}_{ef}%
\\
&\quad
+128\hat F^{ab}\cdot F_{bc}H^2_{adef}\mathcal R^{cdef}%
+32\hat F_{ab}\cdot F^{bc}H^2_{ef}\mathcal R^{aef}{}_c%
+8\hat F_{ab}\cdot F^{ab}H^2_{efgh}R^{efgh}\,,%
\end{split}
\end{equation}
\begin{equation}
\begin{split}
k_3^{(H^2\nabla H)}
&=
128\hat F^{ab}\cdot F^{cd}H^2_{aefc}\nabla_bH_d{}^{ef}%
-128\hat F^{ab}\cdot F^{cd}H^2_{aefc}\nabla^eH^f{}_{bd}%
-64\hat F^{ab}\cdot F^{cd}H^2_{acef}\nabla_bH_d{}^{ef}%
\\
&\quad
+64\hat F^{ab}\cdot F^{cd}H_{ac}{}^e\nabla_bH^2_{de}%
-32\hat F_{ab}\cdot F^{cd}H^{abe}\nabla_cH^2_{de}%
+16\hat F^{ab}\cdot F_{cd}H^2_{abef}(\nabla H)^{cdef}%
\\
&\quad
+64\hat F^{ab}\cdot F^{cd}(H^2)_a{}^e(\nabla H)_{becd}%
-32\hat F^{ab}\cdot F_{cd}H_{abg}H^{cef}(\nabla H)^{dg}{}_{ef}%
-\frac83\hat F^{ab}\cdot F^{cd}H_{abc}\nabla_dH^2%
\\
&\quad
-64\hat F_{ab}\cdot F^{bc}H^{aef}H_{cgh}(\nabla H)_{ef}{}^{gh}%
+32\hat F_{ab}\cdot F^{bc}H^2_{ef}\nabla^eH^{fa}{}_c%
+\frac83\hat F^{ab}\cdot F_{bc}H_a{}^{ce}\nabla_eH^2\,,%
\end{split}
\end{equation}
\begin{equation}
\begin{split}
k_3^{(H^4)}
&=
-32\hat F^{ab}\cdot F^{cd}H^2_{aefc}(H^2)_b{}^{ef}{}_d
-96\hat F^{ab}\cdot F^{cd}H^2_{aefc}(H^2)_{bd}{}^{ef}
-48\hat F^{ab}\cdot F^{cd}(H^2)_{ab}{}^{ef}H^2_{cefd}
\\
&\quad
+32\hat F^{ab}\cdot F^{cd}H^2_{acde}(H^2)_b{}^e
+16\hat F^{ab}\cdot F^{cd}H^2_{abde}(H^2)_c{}^e
-4\hat F_{ab}\cdot F_{cd}H^{abe}H^{fcd}H^2_{ef}
\\
&\quad
-8\hat F_{ab}\cdot F_{cd}H^{eac}H^{fbd}H^2_{ef}
-96\hat F_{ab}\cdot F^{bc}(H^2)^{adef}H^2_{cefd}
+32\hat F_{ab}\cdot F^{bc}(H^2)^{aef}{}_cH^2_{ef}
\\
&\quad
-16\hat F_{ab}\cdot F^{bc}(H^2)^{ad}H^2_{cd}
-2\hat F_{ab}\cdot F^{ab}H^2_{cd}(H^2)^{cd}
\end{split}
\end{equation}
and
\begin{equation}
\begin{split}
k_3^{(\mathrm{e.o.m.})}
&=
32\nabla_e\hat{\mathbbm A}_b\cdot F_{cd}\mathcal R^{cdbe}
-32\hat{\mathbbm A}_b\cdot\nabla^{(-)}_eF_{cd}\mathcal R^{cdbe}
+32\hat{\mathbbm A}_b\cdot F^{cd}H_{cef}\nabla^{(-)}_dH^{bef}
\\
&\quad
-\frac{16}{3}\hat{\mathbbm A}_b\cdot F^{bc}\nabla_cH^2
-32\hat F_{ab}\cdot F_{cd}\mathcal R^{ceab}\mathbbm G^d{}_e
-32\hat F^{ab}\cdot F^{cd}H^2_{abce}(\mathbbm G^e{}_d-\frac12\mathbbm B^e{}_d)
\\
&\quad
-16\hat F_{ab}\cdot F_{cd}H^{eab}H^{fcd}(\mathbbm G_{ef}+\frac12\mathbbm B_{ef})
-128\hat F^{ab}\cdot\nabla^{(-)e}F_{bc}\nabla_{[a}(\mathbbm G^c{}_{e]}+\frac12\mathbbm B^c{}_{e]})
\\
&\quad
-64\hat F^{ab}\cdot\nabla^{(-)e}F_{bc}H^c{}_{f[a}(\mathbbm G^f{}_{e]}+\frac12\mathbbm B^f{}_{e]})
+64\hat F_{ab}\cdot F^{bc}(H^2)^a{}_{efc}\mathbbm G^{ef}
\\
&\quad
-32\hat F_{ab}\cdot F^{bc}H^2_{ce}(\mathbbm G^{ea}+\frac12\mathbbm B^{ea})
-8\hat F_{ab}\cdot F^{ab}H^2_{cd}\mathbbm G^{cd}
+32F_{ab}\cdot\hat F^{cd}H^{eab}\nabla_c\mathbbm B_{ed}
\\
&\quad
-16F^{ab}\cdot\hat F^{cd}H_{eab}\nabla^e\mathbbm B_{cd}
-16\hat F_{ab}\cdot F_{cd}\mathcal R^{ceab}\mathbbm B^d{}_e
+32\hat F_{ab}\cdot F^{cd}\nabla^{(-)}_cH^{eab}\mathbbm B_{de}
\\
&\quad
+64\hat F_{ab}\cdot F^{bc}H^{aef}\nabla_e\mathbbm B_{cf}
-32\hat F_{ab}\cdot F^{bc}H_{cef}\nabla^a\mathbbm B^{ef}
+32\hat F_{ab}\cdot F^{bc}\nabla^aH_{cef}\mathbbm B^{ef}
\\
&\quad
+64\hat F_{ab}\cdot F^{bc}(H^2)^a{}_{efc}\mathbbm B^{ef}
+16\hat F_{ab}\cdot F^{bc}(H^2)^a{}_{cef}\mathbbm B^{ef}
-8\hat F_{ab}\cdot F^{ab}H^{cef}\nabla_c\mathbbm B_{ef}\,.
\end{split}
\label{eq:ell3-eom}
\end{equation}

\section{Reduction of other terms}
The terms defined in (\ref{eq:l1-6}) give rise to the following $O(d,d)$ violating terms quadratic in the KK vectors upon dimensional reduction
\begin{equation}
\begin{split}
l_1
&\rightarrow
\hat F_{ab}\cdot F_{cd}(\nabla H)^{acef}(\nabla H)^{bd}{}_{ef}
+2\hat F_{ab}\cdot\nabla^{(-)e}F_{cd}H^a{}_{ef}(\nabla H)^{bfcd}
\\
&\quad
+\frac12\nabla^{(+)}_e\hat F^{ab}\cdot\nabla^{(-)e}F^{cd}H^2_{acbd}
+\hat F^{ab}\cdot F_{cd}H^2_{aefb}(\nabla H)^{efcd}
+(H\rightarrow-H,\, F\leftrightarrow\hat F)\,,
\\
l_2
&\rightarrow
\hat F_{ab}\cdot F_{cd}(\nabla H)^{abef}(\nabla H)^{cd}{}_{ef}
+2\hat F^{ab}\cdot\nabla^{(-)}_aF_{cd}H_{bef}(\nabla H)^{cdef}
\\
&\quad
+\frac12\nabla^{(+)}_e\hat F^{ab}\cdot\nabla^{(-)e}F^{cd}H^2_{abcd}
+\hat F^{ab}\cdot F_{cd}H^2_{abef}(\nabla H)^{cdef}
+(H\rightarrow-H,\, F\leftrightarrow\hat F)\,,
\\
l_3
&\rightarrow
-4\nabla_d(\hat F_{ab}\cdot F^{bc})\nabla^d(H^2)^a{}_c
+2\nabla_e\hat F^{ab}\cdot\nabla^eF^{cd}H^2_{abcd}
+4\hat F_{ab}\cdot\nabla_eF^{cd}H_{fcd}\nabla^eH^{fab}
\\
&\quad
+2\hat F_{ab}\cdot F^{cd}\nabla^eH^{fab}\nabla_eH_{fcd}
-2\hat F^{ab}\cdot F_{cd}H^{ecd}\nabla_aH^2_{be}%
+2\hat F^{ab}\cdot\nabla_a(F_{cd}H^{ecd})H^2_{be}%
\\
&\quad
+\frac12\hat F^{ab}\cdot F^{cd}H^2_{ac}H^2_{bd}
+\hat F^{ab}\cdot F_{bc}H^2_{ae}(H^2)^{ce}%
+(H\rightarrow-H,\, F\leftrightarrow\hat F)\,,
\\
l_4
&\rightarrow
\hat F_{ab}\cdot F^{cd}\nabla^eH^{fab}\nabla_eH_{fcd}
+\nabla_e\hat F^{ab}\cdot\nabla^eF^{cd}H^2_{abcd}
-4\hat F_{ab}\cdot\nabla_dF^{bc}H^{aef}\nabla^dH_{cef}
\\
&\quad
+4\hat F^{ab}\cdot F^{cd}H^2_{acef}\nabla_bH_d{}^{ef}%
-2\hat F^{ab}\cdot F^{cd}H_{ac}{}^e\nabla_bH^2_{de}%
+\hat F^{ab}\cdot F_{cd}H^2_{abef}(\nabla H)^{cdef}%
\\
&\quad
-\hat F^{ab}\cdot\nabla_bF_{cd}H^2_{ae}H^{ecd}%
-\hat F^{ab}\cdot F_{cd}H^2_{aefb}(H^2)^{cefd}%
-\hat F^{ab}\cdot F^{cd}H^2_{acde}(H^2)_b{}^e%
\\
&\quad
+\frac12\hat F^{ab}\cdot F_{bc}H^2_{ae}(H^2)^{ce}%
+(H\rightarrow-H,\, F\leftrightarrow\hat F)\,,
\\
l_5
&\rightarrow
3\nabla^c\hat F_{ab}\cdot\nabla^dF^{ab}H^2_{cd}
-2\hat F_{ab}\cdot F^{bc}\nabla^aH^{def}\nabla_cH_{def}
+3\hat F_{ab}\cdot F_{cd}H^{abg}H^c{}_{ef}\nabla_gH^{def}%
\\
&\quad
-3\hat F^{ab}\cdot\nabla^eF^{cd}H_{acd}H^2_{be}%
-3\hat F^{ab}\cdot F^{cd}H^2_{ae}\nabla^eH_{bcd}%
-\frac34\hat F^{ab}\cdot F^{cd}H^2_{ac}H^2_{bd}
\\
&\quad
+(H\rightarrow-H,\, F\leftrightarrow\hat F)\,,
\\
l_6
&\rightarrow
6\nabla_c(\hat F_{ab}\cdot F^{ab})\nabla^cH^2
+(H\rightarrow-H,\, F\leftrightarrow\hat F)\,.
\end{split}
\label{eq:l1-6red}
\end{equation}
The terms defined in (\ref{eq:l7-11}) give rise to the following $O(d,d)$ violating terms quadratic in the KK vectors upon dimensional reduction
\begin{equation}
\begin{split}
l_7
&\rightarrow
-2\hat F^{ab}\cdot F^{cd}H^2_{aefc}\mathcal R_{Sb}{}^{ef}{}_d
+2\hat F^{ab}\cdot F^{cd}H^2_{aefc}\mathcal R_{Sbd}{}^{ef}
+2\hat F^{ab}\cdot\nabla^{(-)e}F^{cd}H^f{}_{bd}H^2_{eacf}
\\
&\quad
-\frac12\hat F^{ab}\cdot F_{cd}H^2_{aefb}(H^2)^{cefd}
+(H\rightarrow-H,\,F\leftrightarrow\hat F)\,,
\\
l_8
&\rightarrow
-2\hat F_{ab}\cdot F_{cd}H^{acg}H^{bef}\mathcal R_S^{dg}{}_{ef}%
-\hat F^{ab}\cdot F^{cd}H^2_{acef}\mathcal R_{Sbd}{}^{ef}%
+\hat F^{ab}\cdot F_{cd}H^2_{aefb}\mathcal R_S^{cdef}%
\\
&\quad
+F^{ab}\cdot\nabla^{(+)}_e\hat F^{cd}H_b{}^{ef}H^2_{afcd}%
-\hat F_{ab}\cdot\nabla^{(-)b}F^{cd}H^{aef}H^2_{cefd}%
+\frac12\hat F^{ab}\cdot F_{cd}H^2_{abef}(H^2)^{cefd}
\\
&\quad
+(H\rightarrow-H,\,F\leftrightarrow\hat F)\,,
\\
l_9
&\rightarrow
\hat F_{ab}\cdot F_{cd}H^{abg}H^{cef}\mathcal R_S^d{}_{gef}
-\hat F_{ab}\cdot F_{cd}(H^2)^a{}_e\mathcal R_S^{becd}
-2\hat F^{ab}\cdot F_{bc}H^2_{adef}\mathcal R_S^{cdef}
\\
&\quad
-\frac12F^{ab}\cdot\nabla^{(+)e}\hat F^{cd}H_{bcd}H^2_{ae}
-\frac12F^{ab}\cdot\nabla^{(+)e}\hat F^{cd}H^f{}_{cd}H^2_{abef}
-\hat F^{ab}\cdot \nabla^{(-)}_bF^{cd}H^e{}_{ac}H^2_{de}
\\
&\quad
+\frac12\hat F^{ab}\cdot F^{cd}H^2_{abce}(H^2)_d{}^e
+(H\rightarrow-H,\,F\leftrightarrow\hat F)\,,
\\
l_{10}
&\rightarrow
2\hat F^{ab}\cdot F_{cd}H^2_{abef}\mathcal R_S^{cdef}
-2\hat F^{ab}\cdot F_{bc}H_{aef}H^{cgh}\mathcal R_S^{ef}{}_{gh}
-2\hat F^{ab}\cdot\nabla^{(-)}_bF_{cd}H^{cde}H^2_{ae}
\\
&\quad
+\frac12\hat F^{ab}\cdot F_{cd}H^2_{abef}(H^2)^{cdef}
+(H\rightarrow-H,\,F\leftrightarrow\hat F)\,,
\\
l_{11}
&\rightarrow
4\hat F_{ab}\cdot F^{bc}H^2_{ef}\mathcal R_S^{aef}{}_c
+2\hat F^{cd}\cdot\nabla^{(-)e}F^{ab}H_{acd}H^2_{be}
+\frac12\hat F^{ab}\cdot F^{cd}H^2_{ac}H^2_{bd}
\\
&\quad
+(H\rightarrow-H,\,F\leftrightarrow\hat F)\,.
\end{split}
\label{eq:l7-11red}
\end{equation}
Finally, the terms defined in (\ref{eq:l12-15}) give rise to the following $O(d,d)$ violating terms quadratic in the KK vectors upon dimensional reduction
\begin{equation}
\begin{split}
l_{12}
&\rightarrow
3\hat F_{ab}\cdot F^{cd}(H^2)^{aefb}H^2_{cefd}
-6\hat F^{ab}\cdot F^{cd}(H^2)_{ac}{}^{ef}H^2_{befd}
+(H\rightarrow-H,\,F\leftrightarrow\hat F)\,,
\\
l_{13}
&\rightarrow
2\hat F^{ab}\cdot F_{bc}H^2_{adef}(H^2)^{cefd}
-2\hat F_{ab}\cdot F^{cd}(H^2)^{abef}H^2_{cefd}
-4\hat F^{ab}\cdot F^{cd}H^2_{acde}(H^2)_b{}^e
\\
&\quad
+\hat F_{ab}\cdot F_{cd}H^{eac}H^{fbd}H^2_{ef}
+(H\rightarrow-H,\,F\leftrightarrow\hat F)\,,
\\
l_{14}
&\rightarrow
\hat F^{ab}\cdot F^{cd}H^2_{ac}H^2_{bd}
+4\hat F_{ab}\cdot F^{bc}(H^2)^{aef}{}_cH^2_{ef}
+4\hat F^{ab}\cdot F^{cd}H^2_{abce}(H^2)_d{}^e
\\
&\quad
+(H\rightarrow-H,\,F\leftrightarrow\hat F)\,,
\\
l_{15}
&\rightarrow
-6\hat F^{ab}\cdot F_{bc}H^2_{ad}(H^2)^{cd}
+3\hat F_{ab}\cdot F_{cd}H^{eab}H^{fcd}H^2_{ef}
+(H\rightarrow-H,\,F\leftrightarrow\hat F)\,.
\end{split}
\label{eq:l12-15red}
\end{equation}

\section{Equation of motion terms}\label{app:eom}
In this appendix we collect results on the terms at order $\alpha'^2$ which are proportional to the lowest order equations of motion, given in (\ref{eq:eom1}) and (\ref{eq:eom2}). These terms are not important for our calculations since they can be canceled by field redefinitions at order $\alpha'^2$, producing yet higher order terms. However, in order to read off the field redefinitions required for the $O(d,d)$ violating terms to vanish these terms are needed. Some terms can be canceled by covariant field redefinitions before dimensional reduction, while others would require non-covariant field redefinitions, as we saw already at order $\alpha'$ (or covariant field redefinitions after dimensional reduction).

The terms proportional to the lowest order equations of motion in (\ref{eq:delta-l1}) are given by
\begin{equation}
\delta l_1^{(\mathrm{e.o.m.})}
=
2\nabla^a(F^{bc}\cdot\hat{\mathbbm A}_c)H^2_{ab}
+g_{1ab}\mathbbm G^{ab}
+b_{1ab}\mathbbm B^{ab}\,,
\label{eq:delta-l1-eom}
\end{equation}
where
\begin{equation}
\begin{split}
g_{1ab}&=
-2(\nabla^c-2\partial^c\Phi)\nabla_c\Delta_{ab}
+2(\nabla^c-2\partial^c\Phi)\nabla_a\Delta_{(bc)}
-6\nabla_c\Delta_{[da]}H_b{}^{cd}
-2\nabla_a\Delta^{cd}H_{bcd}
\\
&\quad
+2\Delta_{[ac]}\mathbbm B^c{}_b
+\frac52\Delta^{cd}H^2_{acbd}
-\frac18\Delta^c{}_cH^2_{ab}
+\frac32\Delta_{(ac)}(H^2)^c{}_b
\end{split}
\end{equation}
and
\begin{equation}
\begin{split}
b_{1ab}&=
(\nabla^c-2\partial^c\Phi)\nabla_c\Delta_{ab}
+2(\nabla^c-2\partial^c\Phi)\nabla_a\Delta_{[bc]}
+\frac12\Delta^{cd}R_{abcd}
-3\nabla_c\Delta_{(ad)}H_b{}^{cd}
\\
&\quad
-\frac14\nabla^c\Delta^d{}_dH_{cab}
-\Delta^{(cd)}\nabla_cH_{dab}
-(\nabla_c-2\partial_c\Phi)\Delta^{(cd)}H_{dab}
-\frac32\Delta_{ac}\mathbbm B^c{}_b
\\
&\quad
+\frac18\Delta^c{}_c\mathbbm B_{ab}
+\hat F_{cd}\cdot F_{ef}H^{ecd}H^f{}_{ab}
-\frac14\Delta^{cd}H^2_{acbd}
-\frac14\Delta^{cd}H^2_{abcd}
-\frac12\Delta_{[ac]}(H^2)^c{}_b
\,,
\end{split}
\end{equation}
with $\Delta$ given in (\ref{eq:Delta}).

The terms proportional to the lowest order equations of motion in (\ref{eq:l22}) are
\begin{equation}
\begin{split}
k_{2,2}^{(\mathrm{e.o.m.})}
&=
-\hat{\mathbbm A}^b\cdot F^{cd}H_{cef}\mathcal R^{ef}{}_{bd}%
-\frac14\hat{\mathbbm A}^b\cdot F^{cd}H_{bef}\mathcal R_{cd}{}^{ef}%
+\frac12\hat{\mathbbm A}^b\cdot F^{cd}H_{bef}(\nabla H)_{cd}{}^{ef}%
\\
&\quad
-F^{ab}\cdot\hat F_{bc}H^{cef}\nabla_e(\mathbbm G_{af}+\frac12\mathbbm B_{af})
-\frac12F^{ab}\cdot\hat F_{bc}(H^2)_{ad}{}^{ec}(\mathbbm G^d{}_e+\frac12\mathbbm B^d{}_e)
\\
&\quad
-2\hat F_{ab}\cdot F^{cd}H^{abe}\nabla_c(\mathbbm G_{ed}+\frac12\mathbbm B_{ed})
-\hat F^{ab}\cdot F^{cd}H^2_{abec}(\mathbbm G^e{}_d+\frac12\mathbbm B^e{}_d)
\\
&\quad
+2\hat F^{ab}\cdot F^{cd}H_{ac}{}^e\nabla_{[e}(\mathbbm G_{|b|d]}+\frac12\mathbbm B_{|b|d]})
+\hat F^{ab}\cdot F^{cd}H_{ac}{}^eH_{bf[e}(\mathbbm G^f_{d]}+\frac12\mathbbm B^f{}_{d]})
\\
&\quad
+\frac12\hat F_{ab}\cdot F_{cd}\mathcal R^{abec}\mathbbm B^d{}_e%
+\frac14\hat F_{ab}\cdot F^{bc}\mathcal R_S^a{}_{cef}\mathbbm B^{ef}\,.%
\end{split}
\label{eq:l22-eom}
\end{equation}

The terms in (\ref{eq:l24}) proportional to the lowest order equations of motion take the form
\begin{equation}
\begin{split}
k_{2,4}^{(\mathrm{e.o.m.})}
&=
\frac12\delta l_1^{(\mathrm{e.o.m.})}
+k_{2,2}^{(\mathrm{e.o.m.})}
-\frac{3}{64}k_3^{(\mathrm{e.o.m.})}
+\frac12\hat F^{ab}\cdot\nabla^c\mathbbm A^dH^2_{acbd}
\\
&\quad
+\frac52\hat F^{ab}\cdot\nabla^c\mathbbm A^dH^2_{abcd}
+\frac54\hat F^{ab}\cdot\nabla^c\mathbbm A_aH^2_{bc}
-\hat{\mathbbm A}_b\cdot\nabla^cF^{db}H^2_{cd}
\\
&\quad
+\hat{\mathbbm A}^b\cdot F_{cd}H_{bef}(\nabla H)^{cdef}
+\frac34\hat{\mathbbm A}^b\cdot F_{cd}H^c{}_{ef}\nabla_bH^{def}
+\frac32\hat{\mathbbm A}^b\cdot F_{cd}H^{cef}\nabla^dH_{bef}
\\
&\quad
+\hat{\mathbbm A}^b\cdot F_{cd}H^2_{be}H^{ecd}
-\frac54\hat F_{ab}\cdot F_{cd}H^{eac}H^{fbd}\mathbbm G_{ef}%
+\frac14\hat F_{ab}\cdot F_{cd}H^{eab}H^{fcd}\mathbbm G_{ef}%
\\
&\quad
+7\hat F_{ab}\cdot F^{bc}(H^2)^{aef}{}_c\mathbbm G_{ef}%
-4\hat F^{ab}\cdot F_{bc}H^2_{ad}\mathbbm G^{cd}
-\frac58\hat F_{ab}\cdot F^{ab}H^2_{cd}\mathbbm G^{cd}
\\
&\quad
-\frac12\hat F_{ab}\cdot F_{cd}H_e{}^{ac}\nabla^b\mathbbm B^{de}
-\frac54\hat F_{ab}\cdot F_{cd}H^{ace}\nabla_e\mathbbm B^{bd}
-\frac52\hat F_{ab}\cdot F^{cd}H^{eab}\nabla_c\mathbbm B_{de}
\\
&\quad
+\frac14\hat F^{ab}\cdot F^{cd}H_{eab}\nabla^e\mathbbm B_{cd}
-\frac12\hat F_{ab}\cdot F_{cd}\nabla^cH^{eab}\mathbbm B_e{}^d
+\frac32\hat F_{ab}\cdot F_{cd}\nabla_eH^{cab}\mathbbm B^{ed}
\\
&\quad
-\frac{21}{4}\hat F_{ab}\cdot F^{bc}H_{cef}\nabla^a\mathbbm B^{ef}
-7\hat F_{ab}\cdot F^{bc}H_{cef}\nabla^e\mathbbm B^{fa}
-\frac58\hat F_{ab}\cdot F^{ab}H_{cef}\nabla^c\mathbbm B^{ef}
\\
&\quad
+\frac12F^{ab}\cdot\nabla_c\hat F_{ab}H^{cef}\mathbbm B_{ef}
+\frac12\hat F^{ab}\cdot F^{cd}H^2_{aebd}\mathbbm B_c{}^e
-\frac12\hat F^{ab}\cdot F^{cd}H^2_{abde}\mathbbm B_c{}^e
\\
&\quad
+\frac14\hat F^{ab}\cdot F^{cd}H^2_{ac}\mathbbm B_{bd}
-\frac12\hat F^{ab}\cdot F^{cd}H_{abe}H_{cdf}\mathbbm B^{ef}
-\frac12\hat F^{ab}\cdot F^{cd}H_{abc}H_{def}\mathbbm B^{ef}\,,
\end{split}
\label{eq:l24-eom}
\end{equation}
with $k_3^{(\mathrm{e.o.m.})}$ given in (\ref{eq:ell3-eom}).

Finally, the terms in (\ref{eq:l25}) proportional to the lowest order equations of motion take the form
\begin{equation}
\begin{split}
k_{2,5}^{(\mathrm{e.o.m.})}
&=
k_{2,4}^{(\mathrm{e.o.m.})}
-\frac{11}{4}\hat{\mathbbm A}_a\cdot F^{cd}H^2_{cefd}H^{aef}
-\frac74\hat{\mathbbm A}^a\cdot F^{cd}H_{ac}{}^eH^2_{de}%
-\frac54\hat{\mathbbm A}^a\cdot F_{cd}H^{cde}H^2_{ae}%
\\
&\quad
+3\hat F^{ab}\cdot F^{cd}H^2_{acde}\mathbbm B_b{}^e%
-\frac94\hat F^{ab}\cdot F^{cd}H^2_{abce}\mathbbm B_d{}^e%
-\frac18\hat F^{ab}\cdot F^{cd}H^2_{bd}\mathbbm B_{ac}%
\\
&\quad
+\frac34\hat F^{ab}\cdot F^{cd}H_{acd}H_{bef}\mathbbm B^{ef}
-\frac14\hat F^{ab}\cdot F^{cd}H_{abe}H_{fcd}\mathbbm B^{ef}%
+\frac72\hat F^{ab}\cdot F_b{}^cH^2_{aefc}\mathbbm B^{ef}
\\
&\quad
+\frac14\hat F^{ab}\cdot F_b{}^cH^2_{acef}\mathbbm B^{ef}
-\frac12\hat F_{ab}\cdot F^{bc}H^2_{cd}\mathbbm B^{da}\,.%
\end{split}
\label{eq:l25-eom}
\end{equation}

\bibliographystyle{nb}
\bibliography{biblio}{}
\end{document}